\documentclass[%
reprint,
superscriptaddress,
amsmath,amssymb,
aps,
prb,
]{revtex4-2}
\usepackage{lineno}

\usepackage{graphicx}% Include figure files
\usepackage{dcolumn}% Align table columns on decimal point
\usepackage{bm}% bold math
\usepackage{subfigure}
\usepackage{float}
\usepackage{upgreek}
\begin{document}

	%\switchlinenumbers
	\preprint{APS/123-QED}
	
	\title{Extending the spin coherence lifetimes of ${}^{167}$Er$^{3+}$$:$Y$_2$SiO$_5$ at subkelvin temperatures}% Force line breaks with \\
	
	\author{Jian-Yin Huang}
	\affiliation{CAS Key Laboratory of Quantum Information, University of Science and Technology of China, Hefei 230026, China}%Lines break automatically or can be forced with \\
	
	\affiliation{%
		CAS Center for Excellence in Quantum Information and Quantum Physics, University of Science and Technology of China, Hefei 230026, China
	}%

    \affiliation{Hefei National Laboratory, University of Science and Technology of China, Hefei 230088, China}%
	\author{Pei-Yun Li}
    \affiliation{CAS Key Laboratory of Quantum Information, University of Science and Technology of China, Hefei 230026, China}%Lines break automatically or can be forced with \\

    \affiliation{%
	CAS Center for Excellence in Quantum Information and Quantum Physics, University of Science and Technology of China, Hefei 230026, China
    }%

    \affiliation{Hefei National Laboratory, University of Science and Technology of China, Hefei 230088, China}%
	\author{Zong-Quan Zhou} \thanks{email:zq\_zhou@ustc.edu.cn}
    \affiliation{CAS Key Laboratory of Quantum Information, University of Science and Technology of China, Hefei 230026, China}%Lines break automatically or can be forced with \\

    \affiliation{%
	CAS Center for Excellence in Quantum Information and Quantum Physics, University of Science and Technology of China, Hefei 230026, China
    }%

    \affiliation{Hefei National Laboratory, University of Science and Technology of China, Hefei 230088, China}%
	\author{Chuan-Feng Li} \thanks{email:cﬂi@ustc.edu.cn
	}
    \affiliation{CAS Key Laboratory of Quantum Information, University of Science and Technology of China, Hefei 230026, China}%Lines break automatically or can be forced with \\

    \affiliation{%
	CAS Center for Excellence in Quantum Information and Quantum Physics, University of Science and Technology of China, Hefei 230026, China
    }%

    \affiliation{Hefei National Laboratory, University of Science and Technology of China, Hefei 230088, China}%
	\author{Guang-Can Guo} 
	
	\affiliation{CAS Key Laboratory of Quantum Information, University of Science and Technology of China, Hefei 230026, China}%Lines break automatically or can be forced with \\
	
	\affiliation{%
		CAS Center for Excellence in Quantum Information and Quantum Physics, University of Science and Technology of China, Hefei 230026, China
	}%
	
	\affiliation{Hefei National Laboratory, University of Science and Technology of China, Hefei 230088, China}%

	\date{\today}% It is always \today, today,

	%\pagewiselinenumbers
	%\linenumbers
	\begin{abstract}

		Er$^{3+}$$:$Y$_2$SiO$_5$  is a material of particular interest due to its suitability for telecom-band quantum memories and quantum transducers interfacing optical communication with quantum computers working in the microwave regime. Extending the coherence lifetimes of the electron spins and the nuclear spins is essential for implementing efficient quantum information processing based on such hybrid electron-nuclear spin systems. The electron spin coherence time of Er$^{3+}$$:$Y$_2$SiO$_5$ is so far limited to several microseconds, and there are significant challenges in optimizing coherence lifetimes simultaneously for both the electron and nuclear spins. Here we perform a pulsed-electron-nuclear-double-resonance investigation for an Er$^{3+}$-doped material at subkelvin temperatures. At the lowest working temperature, the electron spin coherence time reaches 290 $\pm$ 17 $\upmu$s, which has been enhanced by 40 times compared with the previous results. In the subkelvin regime, a rapid increase in the nuclear spin coherence time is observed, and the longest coherence time of 738 $\pm$ 6 $\upmu$s is obtained. These extended coherence lifetimes could be valuable resources for further applications of Er$^{3+}$$:$Y$_2$SiO$_5$ in fiber-based quantum networks.
		
	\end{abstract}
	
	\maketitle
	
	%\tableofcontents

	\section{INTRODUCTION}
    The construction of a large-scale quantum network is one of the core topics for quantum information science and technology \cite{gisin2007quantum, kimble2008quantum}. Currently relying on a terrestrial optical fiber network, entanglement distribution can be achieved at a physical separation on the order of 100 km \cite{sangouard2011quantum}. To further extend the distribution distance, the widely accepted solution is to build up a quantum repeater network based on quantum memories \cite{sangouard2011quantum, lvovsky2009optical, liu2021heralded}. Significant progress has been made towards this ultimate goal. Heralded entanglement between two NV centers, physically separated by 1.3 km, is been established by photonic connection at the wavelength of 637 nm \cite{hensen2015loophole} which exhibits strong attenuation in optical fiber. Longer channel length has been recently achieved using the cold atomic ensemble working at the wavelength of 795 nm and being combined with visible-to-telecom wavelength conversion \cite{yu2020entanglement}. However, in such schemes, the networking scale and the data rate will be severely limited by the conversion efficiency and the additional noise. To solve this problem, the necessary requirement is that the matter-based quantum memory should be directly compatible with telecom wavelength. Er$^{3+}$ has a highly coherent optical transition in the telecom C-band \cite{bottger2009effects} and great efforts have been devoted to develop Er$^{3+}$ based quantum memories, using Er$^{3+}$ ensemble \cite{bottger2009effects, lauritzen2010telecommunication, probst2013anisotropic, williamson2014magneto, o2014interfacing, probst2015microwave, ranvcic2018coherence, car2018selective, welinski2019electron, horvath2019extending, saglamyurek2015quantum, saglamyurek2016multiplexed} or single Er$^{3+}$ coupled with optical cavities \cite{dibos2018atomic,raha2020optical,ulanowski2021spectral, chen2020parallel}.   
	
	Among the Er$^{3+}$-doped materials, Er$^{3+}$$:$Y$_2$SiO$_5$ has attracted particular interest \cite{bottger2009effects, lauritzen2010telecommunication, probst2013anisotropic, williamson2014magneto, o2014interfacing, probst2015microwave, ranvcic2018coherence, car2018selective, dibos2018atomic, welinski2019electron, horvath2019extending, raha2020optical}, because of its longest optical coherence lifetime among all of the rare-earth-ion doped solids \cite{bottger2009effects}. In order to further extend the storage time, the optical excitation can be transferred to spin states in a $\Lambda$-type configuration  \cite{lvovsky2009optical, gundougan2015solid, yang2018multiplexed, ma2021one}. The odd isotope ${}^{167}$Er$^{3+}$ can provide such configuration using the nuclear spin states and the nuclear spin coherence lifetime can reach 1.3 s at a magnetic field of 7 T \cite{ranvcic2018coherence}.  The large electronic magnetic moment of Er$^{3+}$$:$Y$_2$SiO$_5$ can be strongly coupled with microwave photons, showing great promises in the conversion of microwave to optical photons \cite{williamson2014magneto, o2014interfacing, everts2019microwave}, as well as microwave quantum memories  \cite{probst2015microwave, afzelius2013proposal, tkalvcec2014strong}. It therefore offers a possibility to build a unified and versatile quantum interface connecting quantum processing, memory and communication. Recently, the emission of single photons from a single Er$^{3+}$ ion is observed \cite{dibos2018atomic,ulanowski2021spectral}, and single-shot spin readout is realized in Y$_2$SiO$_5$ \cite{raha2020optical}. These works have set up the foundation of long-range entanglement distribution based on single Er$^{3+}$ ions.
	
		\begin{figure*}
		\centering
		\subfigure{
			\label{fig:a} %% label for first subfigure
			\includegraphics[width=0.98\columnwidth]{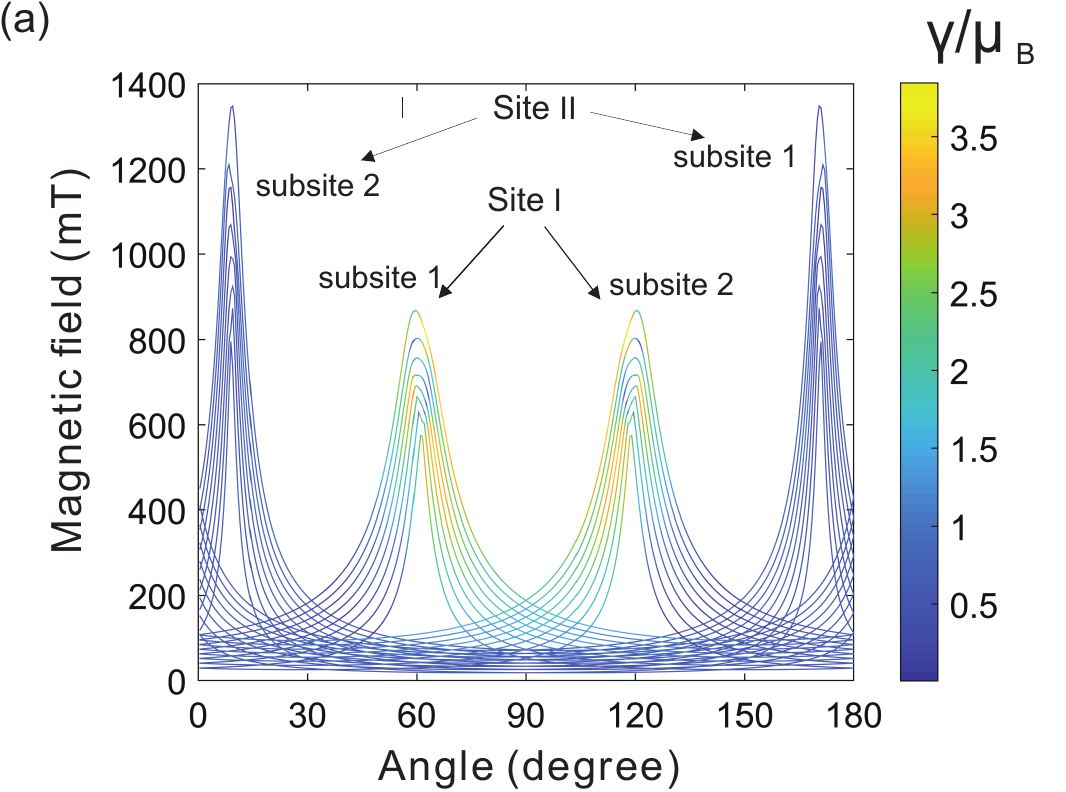}}
		%\hspace{1in}%使第一个子图占一半空间
		\subfigure{
			\label{fig:subfig:b} %% label for secondsubfigure
			\includegraphics[width=1\columnwidth]{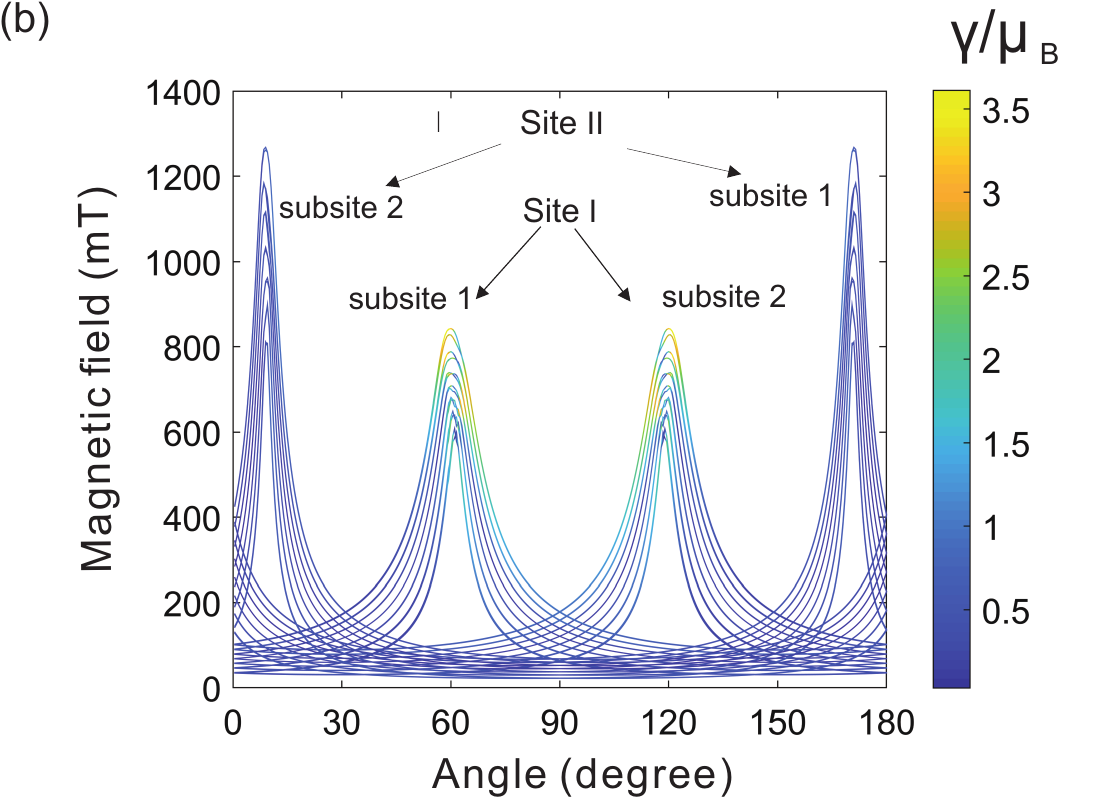}}
		\caption{(color online). (a) Angular variation of the simulated EPR transitions ($\Delta M_I$= 0) of ${}^{167}$Er$^{3+}$ ions at four subsites of Y$_2$SiO$_5$ crystal when the external magnetic field lies in the bD$_1$ plane of the Y$_2$SiO$_5$ crystal.  (b) The simulated spectrum of forbidden transitions ($\Delta M_I$=$\pm$1) varies with the angle of the external field which lies in the bD$_1$ plane. Here the horizontal axes represent the angle with respect to the $b$ axis. Color bars correspond to the transition dipole moment. Signals belonging to the different magnetically inequivalent subsites of the crystallographic sites are marked with arrows.} 
		\label{figb} %% label for entire figure
	\end{figure*}
	The electron and nuclear spin coherence lifetimes are crucial for applications of Er$^{3+}$ ions in  microwave quantum memories \cite{businger2020optical, morton2008solid} and quantum repeater \cite{hensen2015loophole}. However, without utilizing a clock transition \cite{rakonjac2020long} or dynamical decoupling, the electron spin coherence time is so far limited to several microseconds in Y$_2$SiO$_5$ crystals \cite{probst2015microwave, welinski2019electron, raha2020optical}. When the population relaxation lifetime $T_1$ is sufficiently long, the primary decoherence mechanism for this material is the spectral diffusion caused by electron spin flip-flops \cite{bottger2006optical, klauder1962spectral}, which can be inhibited by the polarization of the electron spin ensemble. According to the Boltzmann distribution, electron spin polarization can be achieved by either increasing the magnetic field, or decreasing the sample temperature.
	
	A nuclear spin coherence lifetime of 1.3 s \cite{bottger2009effects} and an optical coherence lifetime of 4 ms \cite{ranvcic2018coherence} has been obtained with working magnetic fields of approximately 7 T. However, there are some drawbacks for using large magnetic fields.  For the single-phonon spin-lattice relaxation process, the electronic population relaxation rate will increase rapidly at the fifth power of the magnetic field \cite{bottger2006optical}. As a result, coherence of the electronic spin has to be sacrificed. Besides, when the magnetic field reaches several teslas, the electronic Zeeman splitting is typically on the order of $\mathcal{O}$(100 GHz), which is far from many promising physical systems for quantum information processing, such as superconducting qubits. Compared with a higher magnetic field, resorting to a lower sample temperature can avoid the problems mentioned above. Moreover, electron spin flips caused by the spin-lattice relaxation (SLR) process will be significantly inhibited due to the reduction in phonon concentration when going to lower temperatures \cite{li2020hyperfine, kukharchyk2019enhancement, bottger2006optical}.

	Pulsed EPR (electron paramagnetic resonance spectroscopy) and ENDOR (electron nuclear double resonance) spectroscopy is the classic methodology to study the coherent dynamics of the coupled electron and nuclear spin systems \cite{morton2008solid, feher1959electron}. Coherence can be transferred between the electronic and nuclear spins with high fidelity \cite{wolfowicz2015coherent}. There have been some previous reports on pulsed-EPR measurements of rare-earth-ion-doped materials at subkelvin temperatures \cite{probst2015microwave, kindem2018characterization}, while pulsed ENDOR spectroscopy at subkelvin temperatures has been implemented only for ${}^{143}$Nd$^{3+}$$:$Y$_2$SiO$_5$ recently \cite{li2020hyperfine}. Here, we implement a pulsed ENDOR investigation for an Er$^{3+}$ doped material at subkelvin temperatures. An electron spin coherence time of 290 $\pm$ 17 $\upmu$s is obtained when the mixing chamber reaches 10 mK, which has been enhanced by 40 times compared to the previous results \cite{probst2015microwave, welinski2019electron, raha2020optical}.  The nuclear spin coherence time increases from 132 to 738 $\upmu$s when the sample temperature drops from 900 to 100 mK, which provides additional subspace for coherence storage.

	\section{EXPERIMENTAL SETUP}

	\begin{figure} 
	\centering
	\includegraphics[width=1\columnwidth]{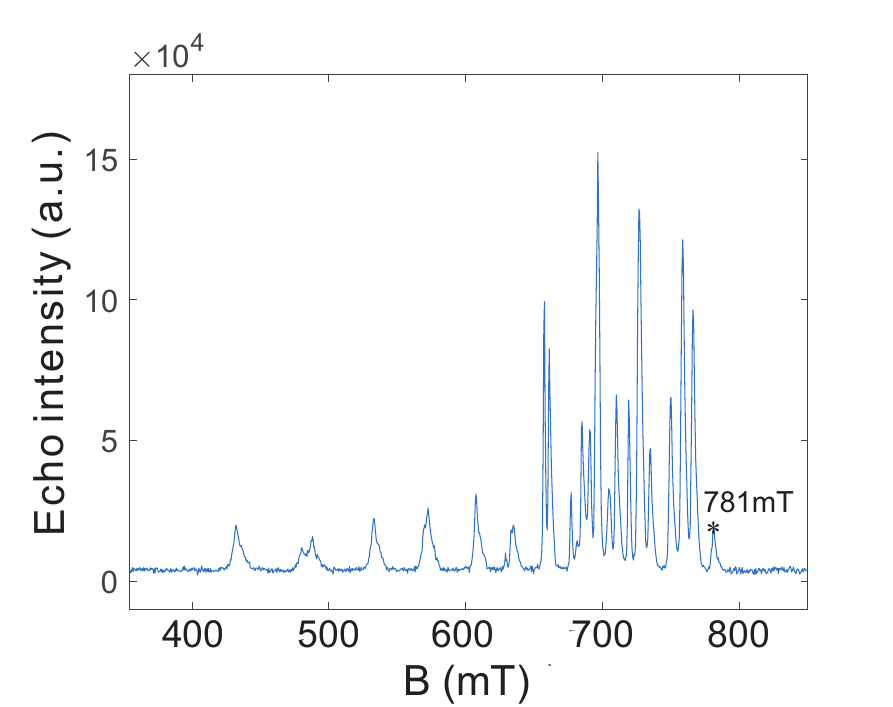}
	
	% some figures do not need to be too wide
	\caption{
		\label{fig:ese}  
		(color online). EPR spectrum of the ${}^{167}$Er$^{3+}$$:$Y$_2$SiO$_5$ detected with field-swept electron spin echo at 4 K. 781 mT is the magnetic field selected for investigations of coherent spin dynamics.}
	\label{figb} %% label for entire figure
    \end{figure}

	Y$_2$SiO$_5$ is a monoclinic crystal belonging to C$_{2h}^{6}$ space group \cite{sun2008magnetic}. It is a widely-adopted host material for rare-earth-based quantum memories, because its constituent elements have small magnetic moments (${}^{89}$Y) or low natural abundance of magnetic isotopes (${}^{29}$Si, $4.6832\%$ abundance, ${}^{17}$O, $0.038\%$ abundance). Long-lived coherence for the substituted rare-earth ions can be expected due to the low spin nature of the host \cite{thiel2011rare}. The Y$_2$SiO$_5$ crystal is doped with 50 ppm Er$^{3+}$ ions and cut along the  D$_1$, D$_2$, $b$ optical extinction axes with dimensions of $1.4\times1\times1.2$ mm$^3$. Er$^{3+}$ is isotopically enriched into ${}^{167}$Er$^{3+}$ with purity of 92\%. ${}^{167}$Er$^{3+}$ has a large nuclear spin quantum number of $I=7/2$. The hyperfine transitions can be utilized as a resource for long-lived quantum memory. In Y$_2$SiO$_5$, Y$^{3+}$ ions are located in two crystallographic sites of C$_1$ symmetry, which can be replaced by Er$^{3+}$ ions. For EPR measurements each of the crystallographic sites can be divided into two magnetically inequivalent subsites related by C$_2$ symmetry around the $b$ axis when the external magnetic ﬁeld is not parallel or perpendicular to the crystal's $b$ axis.  
	
	Traditionally, pulsed ENDOR measurements are hard to apply for general bulk materials at ultralow temperatures due to the large power dissipation required for nuclear spin manipulations \cite{sigillito2017all-electric}. In our recent work we have achieved this goal by successfully combining the pulsed EPR/ENDOR spectrometer within a dilution refrigerator (Triton 400, Oxford Instruments) \cite{li2020hyperfine}.  Here the heating induced by the background noise of the microwave (MW) and radio-frequency (RF) amplifiers is minimized by electrical switches, and the Ohm heating is reduced by using superconducting coaxial cables. The lowest sample temperature is verified to be less than 100 mK. The crystal is placed inside a dielectric ENDOR resonator (Bruker EN4118X-MD4) with the resonance frequency of 9.56 GHz and a $Q$ value of 200. The electromagnet is placed outside the cryostat and the external field is allowed to rotate in a plane. The crystal is mounted in the way such that the external magnetic field $B$ is parallel to the D$_1$b plane of Y$_2$SiO$_5$. Microwave and radio-frequency power used for electron and nuclear spin manipulation is set as 5 and 100 W, respectively.

	\section{ENDOR SPECTROSCOPY}
	For ${}^{167}$Er$^{3+}$$:$Y$_2$SiO$_5$, the spin Hamiltonian can be approximated as \cite{guillot2006hyperfine}:
	
	\begin{equation}
	H=\mu_B  \mathbf{B} \cdot  \mathbf{g} \cdot \mathbf{S} + \mathbf{I \cdot A \cdot S }+ \mathbf{I \cdot Q \cdot I} - \mu_n  g_n \mathbf{ \rm \mathbf{B} \cdot \mathbf{I}},
	\end{equation}
	where $\mu_B$ and $\mu_n$ are the electronic and nuclear Bohr magneton, respectively, $\mathbf{B}$ is the external magnetic field, $\mathbf{g}$ is the Zeeman tensor of Er$^{3+}$, $g_n$ is the nuclear $g$ factor, $\mathbf{Q}$ is the electric quadrupolar tensors and $\mathbf{A}$ is the hyperfine tensors. For ${}^{167}$Er$^{3+}$ with a nuclear spin quantum number of $I=7/2$, there are eight ``allowed" EPR transitions obeying the selection rule of $\Delta M_S= \pm 1$, $\Delta M_I=0$. Here $M_S$ and $M_I$ denote the electron and nuclear spin projections, respectively. However, $M_I$ is typically not a good quantum number due to the low symmetry of Y$_2$SiO$_5$, which results in state mixing among the nuclear spin projections. Therefore, the EPR spectrum should consist of both the $\Delta M_I =0$ and $\Delta M_I= \pm 1$ transitions. When taking into account the $\Delta M_I=\pm 1$ transitions, for a single magnetically inequivalent subsite there should be at most 22 EPR resonance lines observable, including 14 $\Delta M_I=\pm$1 and 8 $\Delta M_I= 0$ transitions. Considering mostly there are two magnetic inequivalent subsites for a single crystallographic site, the dopant Er$^{3+}$ ions locate in four subsites in total. As a result, there should be 88 EPR transitions for an arbitrary field orientation, and the EPR spectrum of ${}^{167}$Er$^{3+}$$:$Y$_2$SiO$_5$ is expected to be very complicated. 
	\begin{figure*}
		\centering
		\subfigure{
			\label{fig: a} %% label for first subfigure
			\includegraphics[width=1\columnwidth]{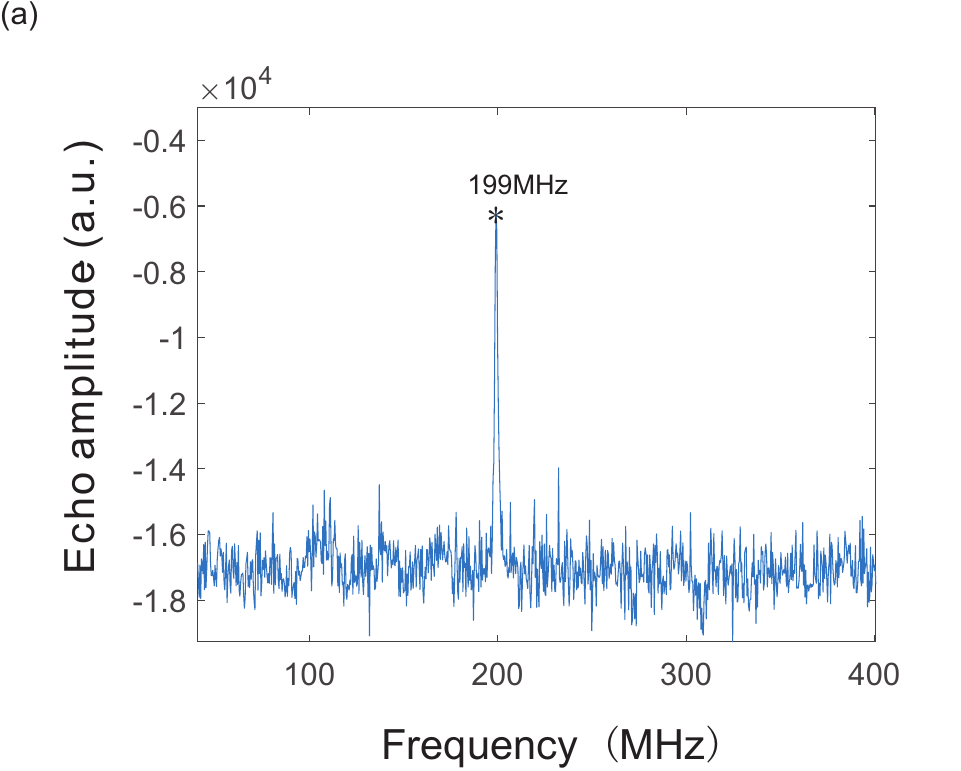}}
		%\hspace{1in}%使第一个子图占一半空间
		\subfigure{
			\label{fig:subfig:b} %% label for secondsubfigure
			\includegraphics[width=1\columnwidth]{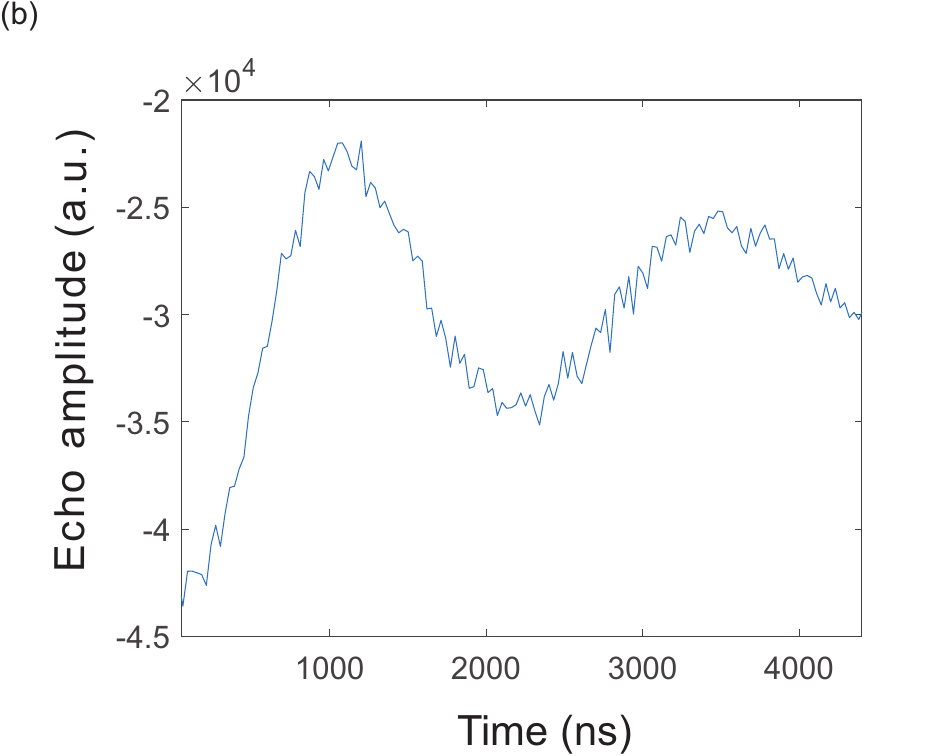}}
		\caption{(color online) (a) Davies ENDOR spectrum recorded at B$_0$ = 781 mT. The NMR resonance frequencies is detected by the Davies ENDOR sequence ($\pi_e-\tau-\pi_{nvarf}-\tau-\frac{\pi}{2}_e-\tau-\pi_e$, where $\pi_{nvarf}$ is the RF pulse that sweeps the frequency), and the resonance peak at 199 MHz corresponds to an NMR transition from $M_I=-7/2$ to $M_I=-5/2$. (b) Rabi oscillation of the NMR transition at 199 MHz. The sequence used to measure the Rabi oscillation is $\pi_e-\tau-\pi_{nvart}-\tau-{\frac{\pi}{2}}_e-\tau-\pi_e$, where $\pi_{nvart}$ is the RF pulse that sweeps the length of pulse. The $\pi$ pulse is determinted to be 1060 ns.}
		\label{figb} %% label for entire figure
	\end{figure*}
	
	EPR signals are gathered with field-swept electron spin echo (ESE) experiments. For most of the field orientations in the bD$_1$ plane, no EPR signal is observed. Instead, the EPR signals occur mainly at four ranges of field orientations with each range spans for approximately 20 $^\circ$ . This phenomenon is consistent with the angular variation of the EPR transition dipole moments $\gamma$=$\vert$$ \langle$$\alpha$$\vert$$\mu_B$$\mathbf{S}$$\cdot$$\mathbf{g}$$\cdot$$\mathbf{\hat{e}}$$_\perp$$\vert$$\beta$$ \rangle$$\vert$. $\alpha$ and $\beta$ are the eigenstates, which are diagonalized from the previously published spin Hamiltonian \cite{guillot2006hyperfine}. $\mathbf{\hat{e}}$$_\perp$ represents the direction of the oscillating magnetic field, which is perpendicular to the external field provided by the electromagnet. The simulated angular dependence of the EPR resonant fields is presented in Fig. 1 along with the corresponding transition strengths. The simulation for all of the four subsites are included. According to the simulation, the large transition dipole moments appear at field orientations characteristic of small effective $g$ factors, which is consistent with previous EPR studies \cite{probst2013anisotropic, abragam2012electron}. In this work, the electromagnet is rotated such that the external field is approximately 57 $^\circ$ with respect to the $b$ axis inside the bD$_1$ plane.

	As shown in Fig. 2, here most of the EPR signals associated with the Er$^{3+}$ spins in subsite 1 of site I can be detected. Moreover, in the current field orientation, the transition intensities of the $\Delta M_I=0$ and $\Delta M_I=\pm 1$ transitions are comparable, which is consistent with the observations in Ref. \cite{guillot2006hyperfine}. When the magnetic field is in the direction corresponding to a small effective $g$ value, the accuracy of the effective spin Hamiltonian is reduced and it is very difficult to simulate the accurate positions of the resonant magnetic fields, or to distinguish between the $\Delta M_I=0$ and $ \Delta M_I=\pm1$ transitions  \cite{guillot2006hyperfine, chen2018hyperfine}. To further study the spin coherent properties, the ``highest field" EPR resonance line at 781 mT in the spectrum is chosen, whose linewidth is 55.8 MHz, which corresponds to the transition between the two electronic spin levels whose energy level spacing is the smallest. It can be readily attributed to the EPR transition corresponding to $M_I=-7/2$, $\Delta M_S=\pm1$, $\Delta M_I=0$. To explore the hyperfine spin transitions, the Davies ENDOR sequence is used to detect the NMR resonance frequencies \cite{schweiger2001principles}. The measured ENDOR signal is given in Fig. 3. The resonance peak at 199 MHz corresponds to the NMR transition from $M_I=-7/2$ to $M_I=-5/2$. The Rabi oscillation of this NMR transition is given in Fig. 3 with a
	$\pi$-pulse determined to be 1060 ns.

	\section{ElECTRON SPIN-LATTICE RELAXATION}
	On the one hand, the electron spin relaxation lifetime ($T_{1e}$) provides a direct upperbound on the electron spin coherence time $T_{2e}$; on the other hand, spin flips of the neighboring electron spins caused by SLR are one of the main sources of decoherence experienced by the central electron and nuclear spins. Based on the obtained EPR transition, we can start the investigations on the coherent spin dynamics. The electronic spin relaxation lifetime $T_{1e}$ of the selected EPR transition is measured with the inversion-recovery sequence ($\rm{\pi-\tau _{var}-\frac{\pi}{2}-\tau_e-\pi-\tau_{e}-echo}$, with varying $\rm{\tau _{var}}$ and fixed $\tau_e$). The length of a MW $\pi$ pulse is determined with Rabi oscillation, which is 52 ns. The recovery curves are fitted with the exponential decay to acquire the $T_{1e}$ data. The temperature dependence of $T_{1e}$ is shown in Fig. 4. For rare-earth ions in solids, population relaxation between the electronic Zeeman levels can be well understood by SLR mechanism  \cite{abragam2012electron}. Typically the SLR mechanism includes the direct process, the Orbach process and the Raman process. The direct process is a one-phonon process, during which resonant phonons are absorbed or released. The Orbach process and Raman process are two-phonon processes \cite{orbach1961spin}. At subkelvin temperatures, the impact of the  two-phonon processes are negligible unless the sample temperature reaches the liquid-helium regime \cite{abragam2012electron}. Therefore, only the direct process is considered when modeling the temperature dependence of $T_{1e}$ in this work. The SLR rate is expressed as \cite{abragam2012electron}: 
	\begin{equation}
	T_{1e}^{-1} = A{\rm coth}(\frac{\Delta E}{2k_BT}),
	\end{equation}
	$A$ is a temperature-independent factor, $\Delta E$ is the energy of the electronic spin transition, $k_B$ is the Boltzmann constant and $T$ is the sample temperature. The relaxation times are given in Fig. 4 with the fitted value of $A$ of 2.91 $\pm0.06$ s$^{-1}$. The data point at 0.1 K is not included when generating the fitted curve. The considerably longer $T_{1e}$ with respect to the value predicted by the fitted curve implies that there may be a phonon bottleneck effect when the temperature goes down to 0.1 K \cite{budoyo2018phonon}.
	\begin{figure} 
		\centering
		\includegraphics[width=1\columnwidth]{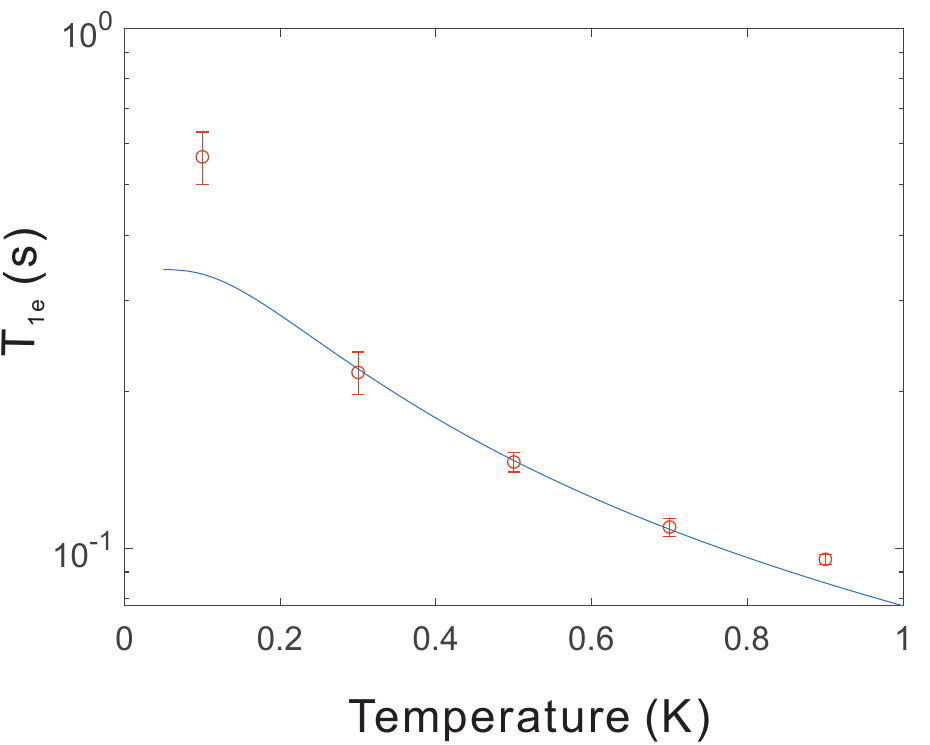}
		
		% some figures do not need to be too wide
		\caption{
			\label{fig:ese}  
			(color online). Temperature dependence of $T_{1e}$ of ${}^{167}$Er$^{3+}$$:$Y$_2$SiO$_5$  at 781 mT. The EPR transition for the measurement of the $T_{1e}$ corresponds to $M_I=-7/2$, $ \Delta M_S=\pm1$, $\Delta M_I=0$. The circles are experimental points with error bars indicating one standard deviations and the solid line is the fitted line based on Eq. $\left(2\right)$.}
		 \label{figb} %% label for entire figure
	\end{figure}
	
	\section{SPIN-SPIN RELAXATION}
	The electron spin coherence time ($T_{2e}$) is measured with Hahn echo sequences ($\rm{{\frac{\pi}{2}}-\tau _{var}-\pi-\tau _{var}-echo}$), as shown in Fig. 5(a). The echo decay curves are fitted using the Mims decay law (echo amplitude E $\propto$ exp[-(2$\tau _{var}/T_{2e})^m]$) \cite{mims1968phase}, and all stretch factors $m$ are between 1 and 1.15 (Fig. 6). The electron spin coherence time increases from 45.1 to 118.6 $\upmu$s with the temperature drops from 900 to 100 mK. 
	
	For ${}^{167}$Er$^{3+}$$:$Y$_2$SiO$_5$, both the spin-lattice relaxation and the spin-spin dipolar interaction can cause fluctuations of the local field, resulting in decoherence of the central spin system.  At sufficiently low temperatures, the spin-lattice interaction is suppressed. In the experimental temperature range, since $T_{1e}$ is on the order of 0.1 s, which is much longer compared with $T_{2e}$, the electronic spin-lattice relaxation does not limit the coherence lifetime. The resonant flip-flops among the dopant electron spins and the host nuclear spins are the primary sources of decoherence in  ${}^{167}$Er$^{3+}$$:$Y$_2$SiO$_5$ \cite{probst2015microwave}. 
	
	The flip-flop process can be divided into the direct process and the indirect process \cite{tyryshkin2012electron}. During the direct process, the central spin directly participates in the flip-flops, while in the indirect process flip-flops among neighboring spins produce the fluctuations of the local magnetic field that dephase the central spin. For the experimental temperature range in this work, the flip-flop of the host nuclear spins and the direct flip-flop among the electron spins are independent of the sample temperature, while the indirect flip-flop process among the electron spins is temperature dependent.  The indirect flip-flop rate is dependent on the number of pairs which can go through the flip-flop process. The number of pairs is proportional to the product of the population residing in each of the electronic Zeeman levels. This can be calculated according to the Boltzmann distribution. Therefore the decoherence rate can be given as \cite{kutter1995electron, probst2015microwave}:
	\begin{equation}
	\frac{1}{T_2}=\sum_{i=1}^{4} \frac{C}{(1+e^{T_{i}/T})(1+e^{-T_{i}/T})}+D,
	\end{equation}
	$T_i$ denotes the effective Zeeman temperatures of the Er$^{3+}$ subensembles residing in the four subsites belonging to both two crystallographic sites. The effective Zeeman temperature of each subensemble is defined as $T_i$=$g_i$$\mu _B$B/k$_B$. $g_i$ is the specific effective $g$ factor corresponding to each of the magnetically inequivalent subsites. $C$ and $D$ are temperature-independent parameters. where $C$ is related to the dipole-dipole interaction strength and $D$ is the residual relaxation rate, which includes the contributions from the flip-flop of the host nuclear spins and the direct flip-flop process. The effective Zeeman temperature of the spin subensemble showing the EPR signal in the current field orientation is 460 mK, as calculated from the MW resonant frequency of 9.56 GHz. The effective Zeeman temperature of subensembles located in the other three
	subsites is calculated using the previously reported $\mathbf{g}$ tensor \cite{guillot2006hyperfine} along with the magnetic field of 781 mT, which are 5.19 K (magnetic inequivalent site 2 of crystallographic site I), 5.91 K (magnetic inequivalent site 2 of crystallographic site \uppercase\expandafter{\romannumeral 2}), and 7.35 K (magnetic inequivalent site 1 of crystallographic site \uppercase\expandafter{\romannumeral 2}), respectively. 
	
	 Due to the large nuclear spin of $I=7/2$, for ${}^{167}$Er$^{3+}$ in a single subsite, there are many EPR transitions at a certain magnetic field. However, only one of them is resonantly excited during a two-pulse-echo sequence. For this reason, compared with previous work \cite{probst2015microwave}, here we further introduce the flip-flop contribution from the spin subensembles coming from the same subsite as the central electronic spins, in Eq. (3). Therefore, the ${}^{167}$Er$^{3+}$ dopants residing in the magnetically inequivalent subsite 1 of the crystallographic site I are providing not only the central electronic spins that generate the EPR signals, but also the environmental electronic spins that generate the noisy magnetic fluctuations. The fitted values of $C$ and $D$ are $60.4 \pm0.41$  and $7.92 \pm2.35$ ms$^{-1}$, respectively. The temperature dependence of $T_{2e}$ is given in Fig. 5(a) with the fitted curve based on Eq. (3). The corresponding decay curves of $T_{2e}$ are provided in Fig. 6 in the Appendix. It has shown an excellent agreement with the experimental data. Therefore it can be deduced that the electronic flip-flops are indeed the primary source of decoherence for the electronic spin of ${}^{167}$Er$^{3+}$$:$Y$_2$SiO$_5$ at subkelvin temperatures.  Considering the magnitude of the effective Zeeman temperatures of subensembles located in the other three subsites, it is obvious that the spins located in the other three subsites are sufficiently polarized when the temperature is below 1 K. The influence from Er$^{3+}$ residing in the other subsites can be ignored and the flip-flop of the ${}^{167}$Er$^{3+}$ belonging to magnetic inequivalent site 1 of crystallographic site I is the major limiting factor to $T_{2e}$. 
	 \begin{figure}
	 	
	 	\subfigure{
	 		\label{fig:subfig:c} %% label for secondsubfigure
	 		\includegraphics[width=0.95\columnwidth]{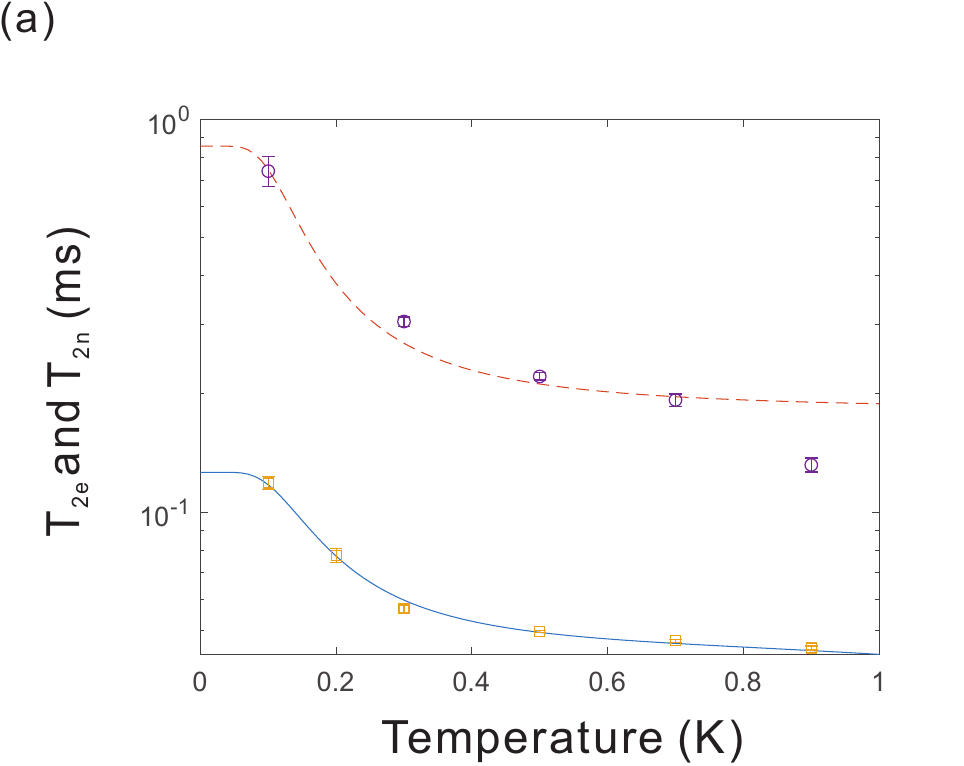}}\\
	 	\subfigure{
	 		\label{fig:subfig:b} %% label for secondsubfigure
	 		\includegraphics[width=1\columnwidth]{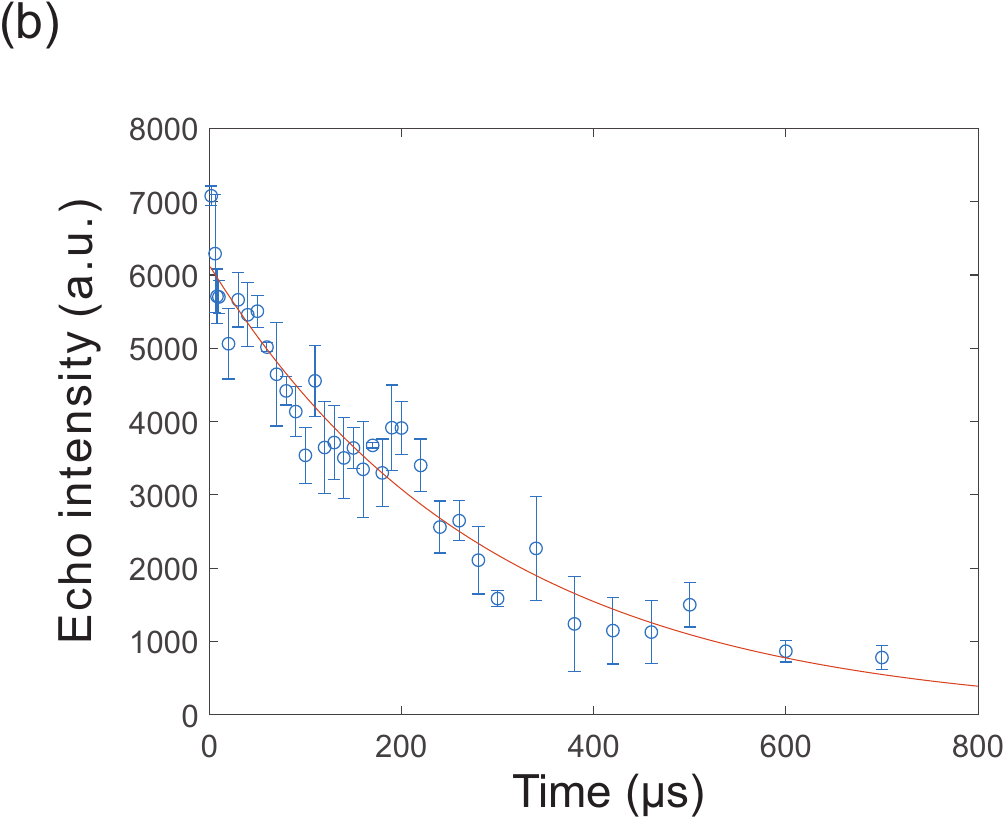}}
	 
	 	%\hspace{1in}%使第一个子图占一半空间
 		\subfigure{
 			\label{fig:a} %% label for first subfigure
 			\includegraphics[width=1\columnwidth]{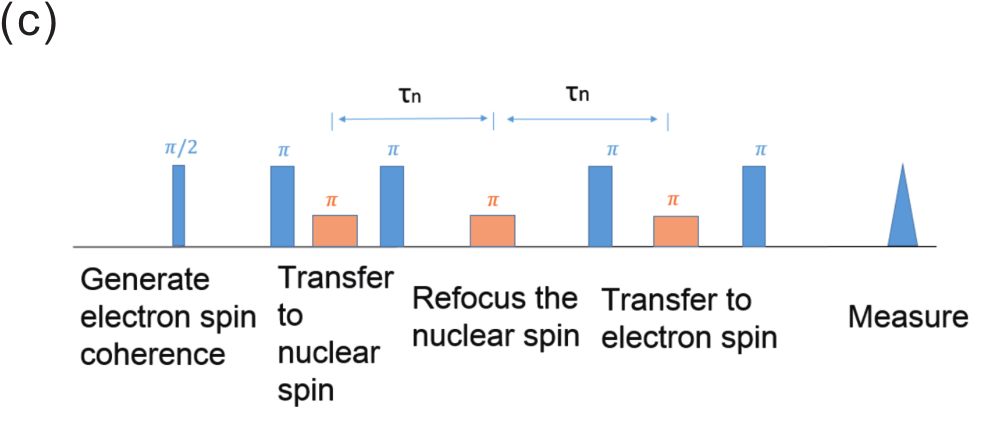}}
	 	\caption{(color online).   (a) Temperature dependence of $T_{2n}$ and $T_{2e}$ in ${}^{167}$Er$^{3+}$$:$Y$_2$SiO$_5$  at 781 mT. The measured $T_{2n}$ and $T_{2e}$ are presented as violet circles and yellow square, respectively. These data are measured by the short and strong $\pi$ pulses (``hard pulse").   The blue solid line and the red dashed line are the fitted curves of the temperature dependencies of the electron and the nuclear spin coherence time, respectively.  (b) The electron spin echo decay at the base temperature. These data are measured by the long and weak pulses (``soft pulse").  (c) Pulse sequence for measuring the nuclear spin coherence lifetime. Microwave pulses and radio frequency pulses, which are employed to manipulate the electron spin and the nuclear spin, are represented by blue and orange filled rectangles, respectively. The electron spin coherence is stored in the nuclear spin for 2$\tau_{n}$ in the sequence where $\tau_n$ is the interval between two RF $\pi$ pulses. }
	 	\label{figb} %% label for entire figure
	 \end{figure}
	 
	 For most measurements in this work, the MW  $\pi$ pulse is short and strong. Under this condition most resonant electronic spins are flipped during the refocusing pulses and the instantaneous spectral diffusion effect tends to be strong \cite{klauder1962spectral}. In order to alleviate the instantaneous spectral diffusion effect, the MW power is reduced to 38 mW and the pulse length is increased to 200 ns. The excitation bandwidth, as well as the number of the excited electronic spins, is thus reduced \cite{dzuba1996selective}. The longest two-pulse-echo electronic spin coherence time is measured as 290 $\pm$ 17 $\upmu$s when the mixing chamber reaches 10 mK, which is considerably longer than the previous result of 7 $\upmu$s \cite{probst2015microwave, welinski2019electron, raha2020optical}. The echo decay curve is given in Fig. 5(b). The significantly extended spin coherence lifetimes indicate that the instantaneous spectral diffusion contributes significantly to decoherence. Slight oscillations can be observed in the echo decay curve, which may be due to the coupling with surrounding nuclear spins. To exclude the possibility of influence from the electron-nuclear spin mixing \cite{rakonjac2020long} for ${}^{167}$Er$^{3+}$, here, a controlled experiment is performed with another piece of crystal containing the even isotope of ${}^{166}$Er$^{3+}$ with zero nuclear spin. The electron spin coherence time of ${}^{166}$Er$^{3+}$$:$Y$_2$SiO$_5$ is measured at 0.7 K with a doping level of 30 ppm. Other experimental conditions, such as the magnetic field orientation, and the length and power of the MW pulses, are kept the same as that for ${}^{167}$Er$^{3+}$$:$Y$_2$SiO$_5$. The measured electron spin coherence time of ${}^{166}$Er$^{3+}$$:$Y$_2$SiO$_5$ is 163 $\pm$3 $\upmu$s (Fig. 8 in the Appendix), which is close to the results measured with  ${}^{167}$Er$^{3+}$$:$Y$_2$SiO$_5$. These results demonstrate that long-lived electron spin coherence has no significant dependence on the type of isotopes. The possible reasons for the different results obtained in our work and in previous works \cite{probst2015microwave, welinski2019electron, raha2020optical} could be the distances between the measured spins and the crystals' surfaces, the thermal conducting properties of the sample holders, and the different coupling strengths with the host nuclei caused by different magnetic field orientations.

   % \begin{figure}[ht] 
    %	\centering
    %	\includegraphics[width=1\columnwidth]{T2_Er166}
    	
    	% some figures do not need to be too wide
    %	\caption{ 
    %		\label{fig:ese}  
    %		(color online). Electron-spin-echo decay of ${}^{166}$Er$^{3+}$:YSO measured at 0.7 K. The circles are experimental points and the blue solid line presents a fit to dates. 
    %	}
    %\end{figure}

	The decoherence time of the NMR transition ($T_{2n}$) is measured by transferring coherence between the NMR and the EPR transitions, with a detailed pulse sequence presented in Fig. 5(c), which is similar to that used in Refs. \cite{morton2008solid, li2020hyperfine}. In this experiment, it is observed that $T_{2n}$ also increases along with the decreasing sample temperature. The corresponding decay curves of $T_{2n}$ are presented in Fig. 7 in the Appendix. The measured coherence time is 738 $\pm$ 6 $\upmu$s at the working temperature of 100 mK. Since magnetic disturbance experienced by the nuclear spin is essentially the same as that of the electron spin, we still use Eq. (3) to model the temperature dependence of $T_{2n}$. The fitted parameters for the nuclear spin coherence in Eq. (3) are $C=17.4 \pm2.2$ ms$^{-1}$, and $D=1.17 \pm0.077$ ms$^{-1}$. As displayed in Fig. 5(a), the fitting based on this model provides a reasonable agreement with the experimental results, which confirms that the temperature dependence of $T_{2n}$ can be well understood with the indirect electronic spin flip-flop processes. The measured ratio between $T_{2n}$ and $T_{2e}$ is not large, due to the fact that at the current magnetic field, the transition frequency of nuclear spin has a relatively large dependence on the field strength, according to the predictions of the spin Hamiltonian \cite{guillot2006hyperfine}.
	
	\section{CONCLUSIONS}
	In conclusion, pulsed EPR and ENDOR spectroscopy of ${}^{167}$Er$^{3+}$$:$Y$_2$SiO$_5$ is investigated at subkelvin temperatures, which is an interesting temperature regime for working with superconductor based quantum computing circuits. Temperature dependence of the electron spin-lattice relaxation time, the electron spin coherence time and the nuclear spin coherence time are characterized from 100 to 900 mK. The measured electron spin relaxation time is shorter than that reported in a similar temperature regime \cite{raha2020optical,probst2013anisotropic,probst2014hybrid}. Such a difference can be well predicted by substituting the different microwave frequencies of these experiments into Eq. (2). Our measurements are performed in a higher magnetic field, and the reduction of electron spin relaxation lifetime is caused by the increased density of phonon modes and the mixing with the higher crystal field levels. The two-pulse spin-echo coherence time of the electron spin is measured to be 290 $\pm$ 17 $\upmu$s when the mixing chamber reaches 10 mK, which has been enhanced by 40 times compared to the previous results  \cite{probst2015microwave, raha2020optical, welinski2019electron}. Meanwhile, our measured electron spin coherence time is slightly longer (but close to) the measured optical coherence lifetime in a similar regime of magnetic field and temperature \cite{kukharchyk2019enhancement, craiciu2021multifunctional}. This result can be expected since the decoherence mechanisms of these two transitions are similar. Besides the application of an ultralow temperature, the electron spin coherence time could be further extended by employing clock-like transitions \cite{rakonjac2020long} or looking for some magnetic field directions with smaller superhyperfine interactions. The coherence time of nuclear spin reaches 738 $\pm$ 6 $\upmu$s and could be further extended with dynamical decoupling \cite{zaripov2013boosting}. Our results suggest that the primary source of decoherence of both the electron and nuclear spins are the flip-flops among the electron spins, which are on the same subsite as the central electron spin.  These results provide a better understanding of the coherent properties of ${}^{167}$Er$^{3+}$$:$Y$_2$SiO$_5$ at subkelvin temperatures. 
	
	\begin{figure}[H]
		\centering
		\includegraphics[width=0.96\columnwidth]{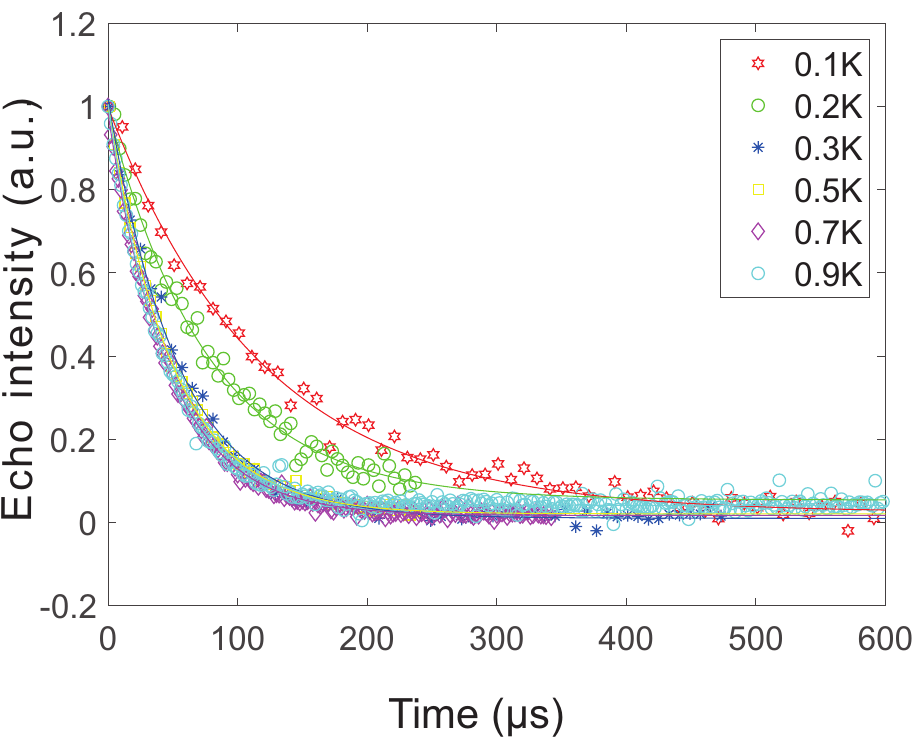}
		
		% some figures do not need to be too wide
		\caption{
			\label{fig:ese}  
			(color online). Electron-spin-echo decay curves of
			${}^{167}$Er$^{3+}$$:$Y$_2$SiO$_5$ measured at different temperatures. The scatter points correspond to the experimental data, and the solid lines with the same color as the scattered points are the fitting curves based on exp[-(2$\tau _{var}/T_{2e})^m]$.}
	\end{figure}
	\begin{acknowledgments}
	This work is supported by the National Key R\&D Program of China (No. 2017YFA0304100), Innovation Program for Quantum Science and Technology (No. 2021ZD0301200), National Natural Science Foundation of China (Nos. 11774331, 11774335, 11821404 and 11654002), and Fundamental Research Funds for the Central Universities (No. WK2470000026 and No. WK2470000029). Z.-Q.Z acknowledges the support from the Youth Innovation Promotion Association CAS.
	\end{acknowledgments}
		
	\section{APPENDIX}
	
	\begin{figure}[ht] 
		\centering
		\includegraphics[width=1\columnwidth]{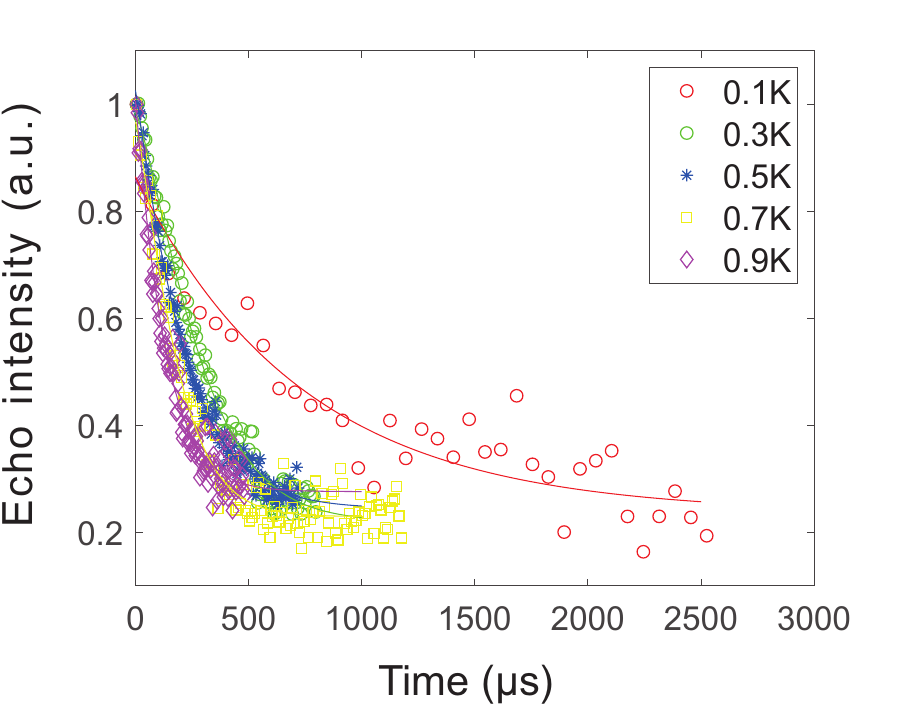}
		
		% some figures do not need to be too wide
		\caption{
			\label{fig:T2e_166}  
			(color online). Nuclear-spin-echo decay curves of
			${}^{167}$Er$^{3+}$$:$Y$_2$SiO$_5$ measured at different temperatures. The scatter points correspond to the experimental data, and the solid lines with the same color as the scattered points are the fitting based on exp[-(2$\tau _{var}/T_{2n})^m]$.}
	\end{figure}
	\begin{figure}[H] 
		\centering
		\includegraphics[width=0.95\columnwidth]{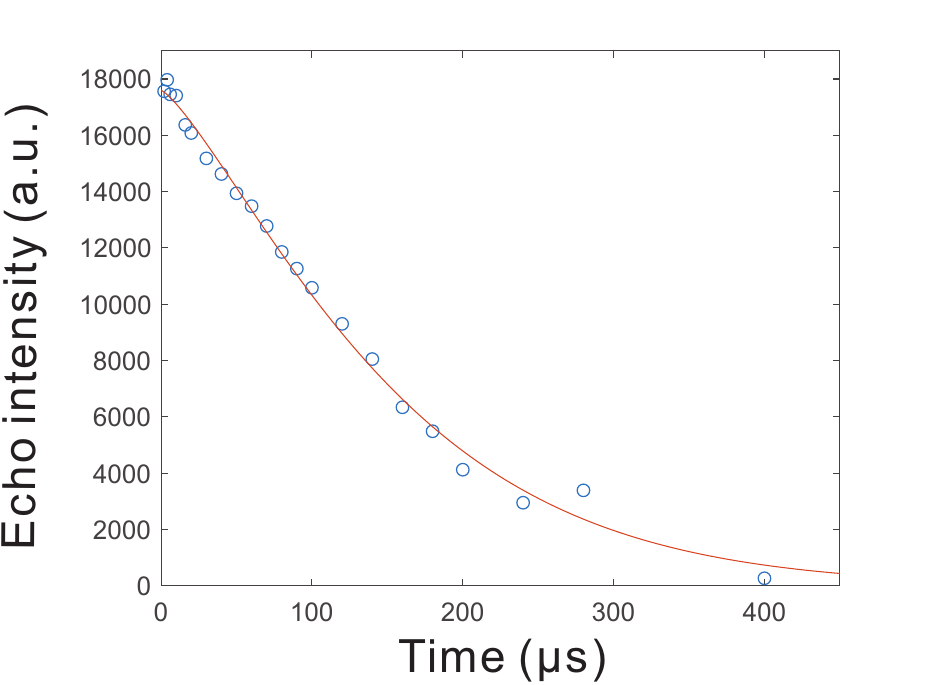}
		
		% some figures do not need to be too wide
		\caption{
			\label{fig:T2e_166}  
			(color online). Electron-spin-echo decay curves of
			${}^{166}$Er$^{3+}$$:$Y$_2$SiO$_5$ measured at 0.7 K. The circles are experimental
			points and the red solid line presents an exponential fit to data.}
	\end{figure}

	 In order to study the indirect flip-flop process of the Er$^{3+}$ electron spin, the coherence lifetimes of the electron spin and the nuclear spin are measured at different temperatures. The echo decay curves for the electron spin and the nuclear spin at various temperatures are provided in Fig. 6 and Fig. 7, respectively. The electron spin echo decay curve of ${}^{166}$Er$^{3+}$$:$Y$_2$SiO$_5$ at a working temperature of 0.7 K is presented in Fig. 8. The doping level of ${}^{166}$Er$^{3+}$ is 30 ppm.

	%\nolinenumbers
	\bibliography{article}% Produces the bibliography via BibTeX.

\begin{thebibliography}{55}%
\makeatletter
\providecommand \@ifxundefined [1]{%
 \@ifx{#1\undefined}
}%
\providecommand \@ifnum [1]{%
 \ifnum #1\expandafter \@firstoftwo
 \else \expandafter \@secondoftwo
 \fi
}%
\providecommand \@ifx [1]{%
 \ifx #1\expandafter \@firstoftwo
 \else \expandafter \@secondoftwo
 \fi
}%
\providecommand \natexlab [1]{#1}%
\providecommand \enquote  [1]{``#1''}%
\providecommand \bibnamefont  [1]{#1}%
\providecommand \bibfnamefont [1]{#1}%
\providecommand \citenamefont [1]{#1}%
\providecommand \href@noop [0]{\@secondoftwo}%
\providecommand \href [0]{\begingroup \@sanitize@url \@href}%
\providecommand \@href[1]{\@@startlink{#1}\@@href}%
\providecommand \@@href[1]{\endgroup#1\@@endlink}%
\providecommand \@sanitize@url [0]{\catcode `\\12\catcode `\$12\catcode
  `\&12\catcode `\#12\catcode `\^12\catcode `\_12\catcode `\%12\relax}%
\providecommand \@@startlink[1]{}%
\providecommand \@@endlink[0]{}%
\providecommand \url  [0]{\begingroup\@sanitize@url \@url }%
\providecommand \@url [1]{\endgroup\@href {#1}{\urlprefix }}%
\providecommand \urlprefix  [0]{URL }%
\providecommand \Eprint [0]{\href }%
\providecommand \doibase [0]{https://doi.org/}%
\providecommand \selectlanguage [0]{\@gobble}%
\providecommand \bibinfo  [0]{\@secondoftwo}%
\providecommand \bibfield  [0]{\@secondoftwo}%
\providecommand \translation [1]{[#1]}%
\providecommand \BibitemOpen [0]{}%
\providecommand \bibitemStop [0]{}%
\providecommand \bibitemNoStop [0]{.\EOS\space}%
\providecommand \EOS [0]{\spacefactor3000\relax}%
\providecommand \BibitemShut  [1]{\csname bibitem#1\endcsname}%
\let\auto@bib@innerbib\@empty
%</preamble>
\bibitem [{\citenamefont {Gisin}\ and\ \citenamefont
  {Thew}(2007)}]{gisin2007quantum}%
  \BibitemOpen
  \bibfield  {author} {\bibinfo {author} {\bibfnamefont {N.}~\bibnamefont
  {Gisin}}\ and\ \bibinfo {author} {\bibfnamefont {R.}~\bibnamefont {Thew}},\
  }\bibfield  {title} {\bibinfo {title} {Quantum communication},\ }\href@noop
  {} {\bibfield  {journal} {\bibinfo  {journal} {Nature photonics}\ }\textbf
  {\bibinfo {volume} {1}},\ \bibinfo {pages} {165} (\bibinfo {year}
  {2007})}\BibitemShut {NoStop}%
\bibitem [{\citenamefont {Kimble}(2008)}]{kimble2008quantum}%
  \BibitemOpen
  \bibfield  {author} {\bibinfo {author} {\bibfnamefont {H.~J.}\ \bibnamefont
  {Kimble}},\ }\bibfield  {title} {\bibinfo {title} {The quantum internet},\
  }\href@noop {} {\bibfield  {journal} {\bibinfo  {journal} {Nature}\ }\textbf
  {\bibinfo {volume} {453}},\ \bibinfo {pages} {1023} (\bibinfo {year}
  {2008})}\BibitemShut {NoStop}%
\bibitem [{\citenamefont {Sangouard}\ \emph {et~al.}(2011)\citenamefont
  {Sangouard}, \citenamefont {Simon}, \citenamefont {De~Riedmatten},\ and\
  \citenamefont {Gisin}}]{sangouard2011quantum}%
  \BibitemOpen
  \bibfield  {author} {\bibinfo {author} {\bibfnamefont {N.}~\bibnamefont
  {Sangouard}}, \bibinfo {author} {\bibfnamefont {C.}~\bibnamefont {Simon}},
  \bibinfo {author} {\bibfnamefont {H.}~\bibnamefont {De~Riedmatten}},\ and\
  \bibinfo {author} {\bibfnamefont {N.}~\bibnamefont {Gisin}},\ }\bibfield
  {title} {\bibinfo {title} {Quantum repeaters based on atomic ensembles and
  linear optics},\ }\href@noop {} {\bibfield  {journal} {\bibinfo  {journal}
  {Reviews of Modern Physics}\ }\textbf {\bibinfo {volume} {83}},\ \bibinfo
  {pages} {33} (\bibinfo {year} {2011})}\BibitemShut {NoStop}%
\bibitem [{\citenamefont {Lvovsky}\ \emph {et~al.}(2009)\citenamefont
  {Lvovsky}, \citenamefont {Sanders},\ and\ \citenamefont
  {Tittel}}]{lvovsky2009optical}%
  \BibitemOpen
  \bibfield  {author} {\bibinfo {author} {\bibfnamefont {A.~I.}\ \bibnamefont
  {Lvovsky}}, \bibinfo {author} {\bibfnamefont {B.~C.}\ \bibnamefont
  {Sanders}},\ and\ \bibinfo {author} {\bibfnamefont {W.}~\bibnamefont
  {Tittel}},\ }\bibfield  {title} {\bibinfo {title} {Optical quantum memory},\
  }\href@noop {} {\bibfield  {journal} {\bibinfo  {journal} {Nature photonics}\
  }\textbf {\bibinfo {volume} {3}},\ \bibinfo {pages} {706} (\bibinfo {year}
  {2009})}\BibitemShut {NoStop}%
\bibitem [{\citenamefont {Liu}\ \emph {et~al.}(2021)\citenamefont {Liu},
  \citenamefont {Hu}, \citenamefont {Li}, \citenamefont {Li}, \citenamefont
  {Li}, \citenamefont {Liang}, \citenamefont {Zhou}, \citenamefont {Li},\ and\
  \citenamefont {Guo}}]{liu2021heralded}%
  \BibitemOpen
  \bibfield  {author} {\bibinfo {author} {\bibfnamefont {X.}~\bibnamefont
  {Liu}}, \bibinfo {author} {\bibfnamefont {J.}~\bibnamefont {Hu}}, \bibinfo
  {author} {\bibfnamefont {Z.-F.}\ \bibnamefont {Li}}, \bibinfo {author}
  {\bibfnamefont {X.}~\bibnamefont {Li}}, \bibinfo {author} {\bibfnamefont
  {P.-Y.}\ \bibnamefont {Li}}, \bibinfo {author} {\bibfnamefont {P.-J.}\
  \bibnamefont {Liang}}, \bibinfo {author} {\bibfnamefont {Z.-Q.}\ \bibnamefont
  {Zhou}}, \bibinfo {author} {\bibfnamefont {C.-F.}\ \bibnamefont {Li}},\ and\
  \bibinfo {author} {\bibfnamefont {G.-C.}\ \bibnamefont {Guo}},\ }\bibfield
  {title} {\bibinfo {title} {Heralded entanglement distribution between two
  absorptive quantum memories},\ }\href@noop {} {\bibfield  {journal} {\bibinfo
   {journal} {Nature}\ }\textbf {\bibinfo {volume} {594}},\ \bibinfo {pages}
  {41} (\bibinfo {year} {2021})}\BibitemShut {NoStop}%
\bibitem [{\citenamefont {Hensen}\ \emph {et~al.}(2015)\citenamefont {Hensen},
  \citenamefont {Bernien}, \citenamefont {Dr{\'e}au}, \citenamefont {Reiserer},
  \citenamefont {Kalb}, \citenamefont {Blok}, \citenamefont {Ruitenberg},
  \citenamefont {Vermeulen}, \citenamefont {Schouten}, \citenamefont
  {Abell{\'a}n} \emph {et~al.}}]{hensen2015loophole}%
  \BibitemOpen
  \bibfield  {author} {\bibinfo {author} {\bibfnamefont {B.}~\bibnamefont
  {Hensen}}, \bibinfo {author} {\bibfnamefont {H.}~\bibnamefont {Bernien}},
  \bibinfo {author} {\bibfnamefont {A.~E.}\ \bibnamefont {Dr{\'e}au}}, \bibinfo
  {author} {\bibfnamefont {A.}~\bibnamefont {Reiserer}}, \bibinfo {author}
  {\bibfnamefont {N.}~\bibnamefont {Kalb}}, \bibinfo {author} {\bibfnamefont
  {M.~S.}\ \bibnamefont {Blok}}, \bibinfo {author} {\bibfnamefont
  {J.}~\bibnamefont {Ruitenberg}}, \bibinfo {author} {\bibfnamefont {R.~F.}\
  \bibnamefont {Vermeulen}}, \bibinfo {author} {\bibfnamefont {R.~N.}\
  \bibnamefont {Schouten}}, \bibinfo {author} {\bibfnamefont {C.}~\bibnamefont
  {Abell{\'a}n}}, \emph {et~al.},\ }\bibfield  {title} {\bibinfo {title}
  {Loophole-free bell inequality violation using electron spins separated by
  1.3 kilometres},\ }\href@noop {} {\bibfield  {journal} {\bibinfo  {journal}
  {Nature}\ }\textbf {\bibinfo {volume} {526}},\ \bibinfo {pages} {682}
  (\bibinfo {year} {2015})}\BibitemShut {NoStop}%
\bibitem [{\citenamefont {Yu}\ \emph {et~al.}(2020)\citenamefont {Yu},
  \citenamefont {Ma}, \citenamefont {Luo}, \citenamefont {Jing}, \citenamefont
  {Sun}, \citenamefont {Fang}, \citenamefont {Yang}, \citenamefont {Liu},
  \citenamefont {Zheng}, \citenamefont {Xie} \emph
  {et~al.}}]{yu2020entanglement}%
  \BibitemOpen
  \bibfield  {author} {\bibinfo {author} {\bibfnamefont {Y.}~\bibnamefont
  {Yu}}, \bibinfo {author} {\bibfnamefont {F.}~\bibnamefont {Ma}}, \bibinfo
  {author} {\bibfnamefont {X.-Y.}\ \bibnamefont {Luo}}, \bibinfo {author}
  {\bibfnamefont {B.}~\bibnamefont {Jing}}, \bibinfo {author} {\bibfnamefont
  {P.-F.}\ \bibnamefont {Sun}}, \bibinfo {author} {\bibfnamefont {R.-Z.}\
  \bibnamefont {Fang}}, \bibinfo {author} {\bibfnamefont {C.-W.}\ \bibnamefont
  {Yang}}, \bibinfo {author} {\bibfnamefont {H.}~\bibnamefont {Liu}}, \bibinfo
  {author} {\bibfnamefont {M.-Y.}\ \bibnamefont {Zheng}}, \bibinfo {author}
  {\bibfnamefont {X.-P.}\ \bibnamefont {Xie}}, \emph {et~al.},\ }\bibfield
  {title} {\bibinfo {title} {Entanglement of two quantum memories via fibres
  over dozens of kilometres},\ }\href@noop {} {\bibfield  {journal} {\bibinfo
  {journal} {Nature}\ }\textbf {\bibinfo {volume} {578}},\ \bibinfo {pages}
  {240} (\bibinfo {year} {2020})}\BibitemShut {NoStop}%
\bibitem [{\citenamefont {B{\"o}ttger}\ \emph {et~al.}(2009)\citenamefont
  {B{\"o}ttger}, \citenamefont {Thiel}, \citenamefont {Cone},\ and\
  \citenamefont {Sun}}]{bottger2009effects}%
  \BibitemOpen
  \bibfield  {author} {\bibinfo {author} {\bibfnamefont {T.}~\bibnamefont
  {B{\"o}ttger}}, \bibinfo {author} {\bibfnamefont {C.}~\bibnamefont {Thiel}},
  \bibinfo {author} {\bibfnamefont {R.}~\bibnamefont {Cone}},\ and\ \bibinfo
  {author} {\bibfnamefont {Y.}~\bibnamefont {Sun}},\ }\bibfield  {title}
  {\bibinfo {title} {{Effects of magnetic field orientation on optical
  decoherence in Er$^{3+}:$Y$_{2}$SiO$_{5}$}},\ }\href@noop {} {\bibfield
  {journal} {\bibinfo  {journal} {Physical Review B}\ }\textbf {\bibinfo
  {volume} {79}},\ \bibinfo {pages} {115104} (\bibinfo {year}
  {2009})}\BibitemShut {NoStop}%
\bibitem [{\citenamefont {Lauritzen}\ \emph {et~al.}(2010)\citenamefont
  {Lauritzen}, \citenamefont {Min{\'a}{\v{r}}}, \citenamefont {De~Riedmatten},
  \citenamefont {Afzelius}, \citenamefont {Sangouard}, \citenamefont {Simon},\
  and\ \citenamefont {Gisin}}]{lauritzen2010telecommunication}%
  \BibitemOpen
  \bibfield  {author} {\bibinfo {author} {\bibfnamefont {B.}~\bibnamefont
  {Lauritzen}}, \bibinfo {author} {\bibfnamefont {J.}~\bibnamefont
  {Min{\'a}{\v{r}}}}, \bibinfo {author} {\bibfnamefont {H.}~\bibnamefont
  {De~Riedmatten}}, \bibinfo {author} {\bibfnamefont {M.}~\bibnamefont
  {Afzelius}}, \bibinfo {author} {\bibfnamefont {N.}~\bibnamefont {Sangouard}},
  \bibinfo {author} {\bibfnamefont {C.}~\bibnamefont {Simon}},\ and\ \bibinfo
  {author} {\bibfnamefont {N.}~\bibnamefont {Gisin}},\ }\bibfield  {title}
  {\bibinfo {title} {Telecommunication-wavelength solid-state memory at the
  single photon level},\ }\href@noop {} {\bibfield  {journal} {\bibinfo
  {journal} {Physical Review Letters}\ }\textbf {\bibinfo {volume} {104}},\
  \bibinfo {pages} {080502} (\bibinfo {year} {2010})}\BibitemShut {NoStop}%
\bibitem [{\citenamefont {Probst}\ \emph {et~al.}(2013)\citenamefont {Probst},
  \citenamefont {Rotzinger}, \citenamefont {W{\"u}nsch}, \citenamefont {Jung},
  \citenamefont {Jerger}, \citenamefont {Siegel}, \citenamefont {Ustinov},\
  and\ \citenamefont {Bushev}}]{probst2013anisotropic}%
  \BibitemOpen
  \bibfield  {author} {\bibinfo {author} {\bibfnamefont {S.}~\bibnamefont
  {Probst}}, \bibinfo {author} {\bibfnamefont {H.}~\bibnamefont {Rotzinger}},
  \bibinfo {author} {\bibfnamefont {S.}~\bibnamefont {W{\"u}nsch}}, \bibinfo
  {author} {\bibfnamefont {P.}~\bibnamefont {Jung}}, \bibinfo {author}
  {\bibfnamefont {M.}~\bibnamefont {Jerger}}, \bibinfo {author} {\bibfnamefont
  {M.}~\bibnamefont {Siegel}}, \bibinfo {author} {\bibfnamefont
  {A.}~\bibnamefont {Ustinov}},\ and\ \bibinfo {author} {\bibfnamefont
  {P.}~\bibnamefont {Bushev}},\ }\bibfield  {title} {\bibinfo {title}
  {Anisotropic rare-earth spin ensemble strongly coupled to a superconducting
  resonator},\ }\href@noop {} {\bibfield  {journal} {\bibinfo  {journal}
  {Physical Review Letters}\ }\textbf {\bibinfo {volume} {110}},\ \bibinfo
  {pages} {157001} (\bibinfo {year} {2013})}\BibitemShut {NoStop}%
\bibitem [{\citenamefont {Williamson}\ \emph {et~al.}(2014)\citenamefont
  {Williamson}, \citenamefont {Chen},\ and\ \citenamefont
  {Longdell}}]{williamson2014magneto}%
  \BibitemOpen
  \bibfield  {author} {\bibinfo {author} {\bibfnamefont {L.~A.}\ \bibnamefont
  {Williamson}}, \bibinfo {author} {\bibfnamefont {Y.-H.}\ \bibnamefont
  {Chen}},\ and\ \bibinfo {author} {\bibfnamefont {J.~J.}\ \bibnamefont
  {Longdell}},\ }\bibfield  {title} {\bibinfo {title} {Magneto-optic modulator
  with unit quantum efficiency},\ }\href@noop {} {\bibfield  {journal}
  {\bibinfo  {journal} {Physical Review Letters}\ }\textbf {\bibinfo {volume}
  {113}},\ \bibinfo {pages} {203601} (\bibinfo {year} {2014})}\BibitemShut
  {NoStop}%
\bibitem [{\citenamefont {O’Brien}\ \emph {et~al.}(2014)\citenamefont
  {O’Brien}, \citenamefont {Lauk}, \citenamefont {Blum}, \citenamefont
  {Morigi},\ and\ \citenamefont {Fleischhauer}}]{o2014interfacing}%
  \BibitemOpen
  \bibfield  {author} {\bibinfo {author} {\bibfnamefont {C.}~\bibnamefont
  {O’Brien}}, \bibinfo {author} {\bibfnamefont {N.}~\bibnamefont {Lauk}},
  \bibinfo {author} {\bibfnamefont {S.}~\bibnamefont {Blum}}, \bibinfo {author}
  {\bibfnamefont {G.}~\bibnamefont {Morigi}},\ and\ \bibinfo {author}
  {\bibfnamefont {M.}~\bibnamefont {Fleischhauer}},\ }\bibfield  {title}
  {\bibinfo {title} {Interfacing superconducting qubits and telecom photons via
  a rare-earth-doped crystal},\ }\href@noop {} {\bibfield  {journal} {\bibinfo
  {journal} {Physical Review Letters}\ }\textbf {\bibinfo {volume} {113}},\
  \bibinfo {pages} {063603} (\bibinfo {year} {2014})}\BibitemShut {NoStop}%
\bibitem [{\citenamefont {Probst}\ \emph {et~al.}(2015)\citenamefont {Probst},
  \citenamefont {Rotzinger}, \citenamefont {Ustinov},\ and\ \citenamefont
  {Bushev}}]{probst2015microwave}%
  \BibitemOpen
  \bibfield  {author} {\bibinfo {author} {\bibfnamefont {S.}~\bibnamefont
  {Probst}}, \bibinfo {author} {\bibfnamefont {H.}~\bibnamefont {Rotzinger}},
  \bibinfo {author} {\bibfnamefont {A.}~\bibnamefont {Ustinov}},\ and\ \bibinfo
  {author} {\bibfnamefont {P.}~\bibnamefont {Bushev}},\ }\bibfield  {title}
  {\bibinfo {title} {Microwave multimode memory with an erbium spin ensemble},\
  }\href@noop {} {\bibfield  {journal} {\bibinfo  {journal} {Physical Review
  B}\ }\textbf {\bibinfo {volume} {92}},\ \bibinfo {pages} {014421} (\bibinfo
  {year} {2015})}\BibitemShut {NoStop}%
\bibitem [{\citenamefont {Ran{\v{c}}i{\'c}}\ \emph {et~al.}(2018)\citenamefont
  {Ran{\v{c}}i{\'c}}, \citenamefont {Hedges}, \citenamefont {Ahlefeldt},\ and\
  \citenamefont {Sellars}}]{ranvcic2018coherence}%
  \BibitemOpen
  \bibfield  {author} {\bibinfo {author} {\bibfnamefont {M.}~\bibnamefont
  {Ran{\v{c}}i{\'c}}}, \bibinfo {author} {\bibfnamefont {M.~P.}\ \bibnamefont
  {Hedges}}, \bibinfo {author} {\bibfnamefont {R.~L.}\ \bibnamefont
  {Ahlefeldt}},\ and\ \bibinfo {author} {\bibfnamefont {M.~J.}\ \bibnamefont
  {Sellars}},\ }\bibfield  {title} {\bibinfo {title} {Coherence time of over a
  second in a telecom-compatible quantum memory storage material},\ }\href@noop
  {} {\bibfield  {journal} {\bibinfo  {journal} {Nature Physics}\ }\textbf
  {\bibinfo {volume} {14}},\ \bibinfo {pages} {50} (\bibinfo {year}
  {2018})}\BibitemShut {NoStop}%
\bibitem [{\citenamefont {Car}\ \emph {et~al.}(2018)\citenamefont {Car},
  \citenamefont {Veissier}, \citenamefont {Louchet-Chauvet}, \citenamefont
  {Le~Gou{\"e}t},\ and\ \citenamefont {Chaneli{\`e}re}}]{car2018selective}%
  \BibitemOpen
  \bibfield  {author} {\bibinfo {author} {\bibfnamefont {B.}~\bibnamefont
  {Car}}, \bibinfo {author} {\bibfnamefont {L.}~\bibnamefont {Veissier}},
  \bibinfo {author} {\bibfnamefont {A.}~\bibnamefont {Louchet-Chauvet}},
  \bibinfo {author} {\bibfnamefont {J.-L.}\ \bibnamefont {Le~Gou{\"e}t}},\ and\
  \bibinfo {author} {\bibfnamefont {T.}~\bibnamefont {Chaneli{\`e}re}},\
  }\bibfield  {title} {\bibinfo {title} {Selective optical addressing of
  nuclear spins through superhyperfine interaction in rare-earth doped
  solids},\ }\href@noop {} {\bibfield  {journal} {\bibinfo  {journal} {Physical
  Review Letters}\ }\textbf {\bibinfo {volume} {120}},\ \bibinfo {pages}
  {197401} (\bibinfo {year} {2018})}\BibitemShut {NoStop}%
\bibitem [{\citenamefont {Welinski}\ \emph {et~al.}(2019)\citenamefont
  {Welinski}, \citenamefont {Woodburn}, \citenamefont {Lauk}, \citenamefont
  {Cone}, \citenamefont {Simon}, \citenamefont {Goldner},\ and\ \citenamefont
  {Thiel}}]{welinski2019electron}%
  \BibitemOpen
  \bibfield  {author} {\bibinfo {author} {\bibfnamefont {S.}~\bibnamefont
  {Welinski}}, \bibinfo {author} {\bibfnamefont {P.~J.}\ \bibnamefont
  {Woodburn}}, \bibinfo {author} {\bibfnamefont {N.}~\bibnamefont {Lauk}},
  \bibinfo {author} {\bibfnamefont {R.~L.}\ \bibnamefont {Cone}}, \bibinfo
  {author} {\bibfnamefont {C.}~\bibnamefont {Simon}}, \bibinfo {author}
  {\bibfnamefont {P.}~\bibnamefont {Goldner}},\ and\ \bibinfo {author}
  {\bibfnamefont {C.~W.}\ \bibnamefont {Thiel}},\ }\bibfield  {title} {\bibinfo
  {title} {Electron spin coherence in optically excited states of rare-earth
  ions for microwave to optical quantum transducers},\ }\href@noop {}
  {\bibfield  {journal} {\bibinfo  {journal} {Physical Review Letters}\
  }\textbf {\bibinfo {volume} {122}},\ \bibinfo {pages} {247401} (\bibinfo
  {year} {2019})}\BibitemShut {NoStop}%
\bibitem [{\citenamefont {Horvath}\ \emph {et~al.}(2019)\citenamefont
  {Horvath}, \citenamefont {Rakonjac}, \citenamefont {Chen}, \citenamefont
  {Longdell}, \citenamefont {Goldner}, \citenamefont {Wells},\ and\
  \citenamefont {Reid}}]{horvath2019extending}%
  \BibitemOpen
  \bibfield  {author} {\bibinfo {author} {\bibfnamefont {S.~P.}\ \bibnamefont
  {Horvath}}, \bibinfo {author} {\bibfnamefont {J.~V.}\ \bibnamefont
  {Rakonjac}}, \bibinfo {author} {\bibfnamefont {Y.-H.}\ \bibnamefont {Chen}},
  \bibinfo {author} {\bibfnamefont {J.~J.}\ \bibnamefont {Longdell}}, \bibinfo
  {author} {\bibfnamefont {P.}~\bibnamefont {Goldner}}, \bibinfo {author}
  {\bibfnamefont {J.-P.}\ \bibnamefont {Wells}},\ and\ \bibinfo {author}
  {\bibfnamefont {M.~F.}\ \bibnamefont {Reid}},\ }\bibfield  {title} {\bibinfo
  {title} {{Extending Phenomenological Crystal-Field Methods to $C_{1}$
  Point-Group Symmetry: Characterization of the Optically Excited Hyperfine
  Structure of ${}^{167}$Er$^{3+}:$Y$_{2}$SiO$_{5}$}},\ }\href@noop {}
  {\bibfield  {journal} {\bibinfo  {journal} {Physical Review Letters}\
  }\textbf {\bibinfo {volume} {123}},\ \bibinfo {pages} {057401} (\bibinfo
  {year} {2019})}\BibitemShut {NoStop}%
\bibitem [{\citenamefont {Saglamyurek}\ \emph {et~al.}(2015)\citenamefont
  {Saglamyurek}, \citenamefont {Jin}, \citenamefont {Verma}, \citenamefont
  {Shaw}, \citenamefont {Marsili}, \citenamefont {Nam}, \citenamefont {Oblak},\
  and\ \citenamefont {Tittel}}]{saglamyurek2015quantum}%
  \BibitemOpen
  \bibfield  {author} {\bibinfo {author} {\bibfnamefont {E.}~\bibnamefont
  {Saglamyurek}}, \bibinfo {author} {\bibfnamefont {J.}~\bibnamefont {Jin}},
  \bibinfo {author} {\bibfnamefont {V.~B.}\ \bibnamefont {Verma}}, \bibinfo
  {author} {\bibfnamefont {M.~D.}\ \bibnamefont {Shaw}}, \bibinfo {author}
  {\bibfnamefont {F.}~\bibnamefont {Marsili}}, \bibinfo {author} {\bibfnamefont
  {S.~W.}\ \bibnamefont {Nam}}, \bibinfo {author} {\bibfnamefont
  {D.}~\bibnamefont {Oblak}},\ and\ \bibinfo {author} {\bibfnamefont
  {W.}~\bibnamefont {Tittel}},\ }\bibfield  {title} {\bibinfo {title} {Quantum
  storage of entangled telecom-wavelength photons in an erbium-doped optical
  fibre},\ }\href@noop {} {\bibfield  {journal} {\bibinfo  {journal} {Nature
  Photonics}\ }\textbf {\bibinfo {volume} {9}},\ \bibinfo {pages} {83}
  (\bibinfo {year} {2015})}\BibitemShut {NoStop}%
\bibitem [{\citenamefont {Saglamyurek}\ \emph {et~al.}(2016)\citenamefont
  {Saglamyurek}, \citenamefont {Puigibert}, \citenamefont {Zhou}, \citenamefont
  {Giner}, \citenamefont {Marsili}, \citenamefont {Verma}, \citenamefont {Nam},
  \citenamefont {Oesterling}, \citenamefont {Nippa}, \citenamefont {Oblak}
  \emph {et~al.}}]{saglamyurek2016multiplexed}%
  \BibitemOpen
  \bibfield  {author} {\bibinfo {author} {\bibfnamefont {E.}~\bibnamefont
  {Saglamyurek}}, \bibinfo {author} {\bibfnamefont {M.~G.}\ \bibnamefont
  {Puigibert}}, \bibinfo {author} {\bibfnamefont {Q.}~\bibnamefont {Zhou}},
  \bibinfo {author} {\bibfnamefont {L.}~\bibnamefont {Giner}}, \bibinfo
  {author} {\bibfnamefont {F.}~\bibnamefont {Marsili}}, \bibinfo {author}
  {\bibfnamefont {V.~B.}\ \bibnamefont {Verma}}, \bibinfo {author}
  {\bibfnamefont {S.~W.}\ \bibnamefont {Nam}}, \bibinfo {author} {\bibfnamefont
  {L.}~\bibnamefont {Oesterling}}, \bibinfo {author} {\bibfnamefont
  {D.}~\bibnamefont {Nippa}}, \bibinfo {author} {\bibfnamefont
  {D.}~\bibnamefont {Oblak}}, \emph {et~al.},\ }\bibfield  {title} {\bibinfo
  {title} {A multiplexed light-matter interface for fibre-based quantum
  networks},\ }\href@noop {} {\bibfield  {journal} {\bibinfo  {journal} {Nature
  communications}\ }\textbf {\bibinfo {volume} {7}},\ \bibinfo {pages} {11202}
  (\bibinfo {year} {2016})}\BibitemShut {NoStop}%
\bibitem [{\citenamefont {Dibos}\ \emph {et~al.}(2018)\citenamefont {Dibos},
  \citenamefont {Raha}, \citenamefont {Phenicie},\ and\ \citenamefont
  {Thompson}}]{dibos2018atomic}%
  \BibitemOpen
  \bibfield  {author} {\bibinfo {author} {\bibfnamefont {A.}~\bibnamefont
  {Dibos}}, \bibinfo {author} {\bibfnamefont {M.}~\bibnamefont {Raha}},
  \bibinfo {author} {\bibfnamefont {C.}~\bibnamefont {Phenicie}},\ and\
  \bibinfo {author} {\bibfnamefont {J.~D.}\ \bibnamefont {Thompson}},\
  }\bibfield  {title} {\bibinfo {title} {Atomic source of single photons in the
  telecom band},\ }\href@noop {} {\bibfield  {journal} {\bibinfo  {journal}
  {Physical Review Letters}\ }\textbf {\bibinfo {volume} {120}},\ \bibinfo
  {pages} {243601} (\bibinfo {year} {2018})}\BibitemShut {NoStop}%
\bibitem [{\citenamefont {Raha}\ \emph {et~al.}(2020)\citenamefont {Raha},
  \citenamefont {Chen}, \citenamefont {Phenicie}, \citenamefont {Ourari},
  \citenamefont {Dibos},\ and\ \citenamefont {Thompson}}]{raha2020optical}%
  \BibitemOpen
  \bibfield  {author} {\bibinfo {author} {\bibfnamefont {M.}~\bibnamefont
  {Raha}}, \bibinfo {author} {\bibfnamefont {S.}~\bibnamefont {Chen}}, \bibinfo
  {author} {\bibfnamefont {C.~M.}\ \bibnamefont {Phenicie}}, \bibinfo {author}
  {\bibfnamefont {S.}~\bibnamefont {Ourari}}, \bibinfo {author} {\bibfnamefont
  {A.~M.}\ \bibnamefont {Dibos}},\ and\ \bibinfo {author} {\bibfnamefont
  {J.~D.}\ \bibnamefont {Thompson}},\ }\bibfield  {title} {\bibinfo {title}
  {Optical quantum nondemolition measurement of a single rare earth ion
  qubit},\ }\href@noop {} {\bibfield  {journal} {\bibinfo  {journal} {Nature
  communications}\ }\textbf {\bibinfo {volume} {11}},\ \bibinfo {pages} {1605}
  (\bibinfo {year} {2020})}\BibitemShut {NoStop}%
\bibitem [{\citenamefont {Ulanowski}\ \emph {et~al.}(2021)\citenamefont
  {Ulanowski}, \citenamefont {Merkel},\ and\ \citenamefont
  {Reiserer}}]{ulanowski2021spectral}%
  \BibitemOpen
  \bibfield  {author} {\bibinfo {author} {\bibfnamefont {A.}~\bibnamefont
  {Ulanowski}}, \bibinfo {author} {\bibfnamefont {B.}~\bibnamefont {Merkel}},\
  and\ \bibinfo {author} {\bibfnamefont {A.}~\bibnamefont {Reiserer}},\
  }\bibfield  {title} {\bibinfo {title} {Spectral multiplexing of telecom
  emitters with stable transition frequency},\ }\href@noop {} {\bibfield
  {journal} {\bibinfo  {journal} {arXiv preprint arXiv:2110.09409}\ } (\bibinfo
  {year} {2021})}\BibitemShut {NoStop}%
\bibitem [{\citenamefont {Chen}\ \emph {et~al.}(2020)\citenamefont {Chen},
  \citenamefont {Raha}, \citenamefont {Phenicie}, \citenamefont {Ourari},\ and\
  \citenamefont {Thompson}}]{chen2020parallel}%
  \BibitemOpen
  \bibfield  {author} {\bibinfo {author} {\bibfnamefont {S.}~\bibnamefont
  {Chen}}, \bibinfo {author} {\bibfnamefont {M.}~\bibnamefont {Raha}}, \bibinfo
  {author} {\bibfnamefont {C.~M.}\ \bibnamefont {Phenicie}}, \bibinfo {author}
  {\bibfnamefont {S.}~\bibnamefont {Ourari}},\ and\ \bibinfo {author}
  {\bibfnamefont {J.~D.}\ \bibnamefont {Thompson}},\ }\bibfield  {title}
  {\bibinfo {title} {Parallel single-shot measurement and coherent control of
  solid-state spins below the diffraction limit},\ }\href@noop {} {\bibfield
  {journal} {\bibinfo  {journal} {Science}\ }\textbf {\bibinfo {volume}
  {370}},\ \bibinfo {pages} {592} (\bibinfo {year} {2020})}\BibitemShut
  {NoStop}%
\bibitem [{\citenamefont {G{\"u}ndo{\u{g}}an}\ \emph
  {et~al.}(2015)\citenamefont {G{\"u}ndo{\u{g}}an}, \citenamefont {Ledingham},
  \citenamefont {Kutluer}, \citenamefont {Mazzera},\ and\ \citenamefont
  {De~Riedmatten}}]{gundougan2015solid}%
  \BibitemOpen
  \bibfield  {author} {\bibinfo {author} {\bibfnamefont {M.}~\bibnamefont
  {G{\"u}ndo{\u{g}}an}}, \bibinfo {author} {\bibfnamefont {P.~M.}\ \bibnamefont
  {Ledingham}}, \bibinfo {author} {\bibfnamefont {K.}~\bibnamefont {Kutluer}},
  \bibinfo {author} {\bibfnamefont {M.}~\bibnamefont {Mazzera}},\ and\ \bibinfo
  {author} {\bibfnamefont {H.}~\bibnamefont {De~Riedmatten}},\ }\bibfield
  {title} {\bibinfo {title} {Solid state spin-wave quantum memory for time-bin
  qubits},\ }\href@noop {} {\bibfield  {journal} {\bibinfo  {journal} {Physical
  review letters}\ }\textbf {\bibinfo {volume} {114}},\ \bibinfo {pages}
  {230501} (\bibinfo {year} {2015})}\BibitemShut {NoStop}%
\bibitem [{\citenamefont {Yang}\ \emph {et~al.}(2018)\citenamefont {Yang},
  \citenamefont {Zhou}, \citenamefont {Hua}, \citenamefont {Liu}, \citenamefont
  {Li}, \citenamefont {Li}, \citenamefont {Ma}, \citenamefont {Liu},
  \citenamefont {Liang}, \citenamefont {Li} \emph
  {et~al.}}]{yang2018multiplexed}%
  \BibitemOpen
  \bibfield  {author} {\bibinfo {author} {\bibfnamefont {T.-S.}\ \bibnamefont
  {Yang}}, \bibinfo {author} {\bibfnamefont {Z.-Q.}\ \bibnamefont {Zhou}},
  \bibinfo {author} {\bibfnamefont {Y.-L.}\ \bibnamefont {Hua}}, \bibinfo
  {author} {\bibfnamefont {X.}~\bibnamefont {Liu}}, \bibinfo {author}
  {\bibfnamefont {Z.-F.}\ \bibnamefont {Li}}, \bibinfo {author} {\bibfnamefont
  {P.-Y.}\ \bibnamefont {Li}}, \bibinfo {author} {\bibfnamefont
  {Y.}~\bibnamefont {Ma}}, \bibinfo {author} {\bibfnamefont {C.}~\bibnamefont
  {Liu}}, \bibinfo {author} {\bibfnamefont {P.-J.}\ \bibnamefont {Liang}},
  \bibinfo {author} {\bibfnamefont {X.}~\bibnamefont {Li}}, \emph {et~al.},\
  }\bibfield  {title} {\bibinfo {title} {Multiplexed storage and real-time
  manipulation based on a multiple degree-of-freedom quantum memory},\
  }\href@noop {} {\bibfield  {journal} {\bibinfo  {journal} {Nature
  communications}\ }\textbf {\bibinfo {volume} {9}},\ \bibinfo {pages} {3407}
  (\bibinfo {year} {2018})}\BibitemShut {NoStop}%
\bibitem [{\citenamefont {Ma}\ \emph {et~al.}(2021)\citenamefont {Ma},
  \citenamefont {Ma}, \citenamefont {Zhou}, \citenamefont {Li},\ and\
  \citenamefont {Guo}}]{ma2021one}%
  \BibitemOpen
  \bibfield  {author} {\bibinfo {author} {\bibfnamefont {Y.}~\bibnamefont
  {Ma}}, \bibinfo {author} {\bibfnamefont {Y.-Z.}\ \bibnamefont {Ma}}, \bibinfo
  {author} {\bibfnamefont {Z.-Q.}\ \bibnamefont {Zhou}}, \bibinfo {author}
  {\bibfnamefont {C.-F.}\ \bibnamefont {Li}},\ and\ \bibinfo {author}
  {\bibfnamefont {G.-C.}\ \bibnamefont {Guo}},\ }\bibfield  {title} {\bibinfo
  {title} {One-hour coherent optical storage in an atomic frequency comb
  memory},\ }\href@noop {} {\bibfield  {journal} {\bibinfo  {journal} {Nature
  communications}\ }\textbf {\bibinfo {volume} {12}},\ \bibinfo {pages} {2381}
  (\bibinfo {year} {2021})}\BibitemShut {NoStop}%
\bibitem [{\citenamefont {Everts}\ \emph {et~al.}(2019)\citenamefont {Everts},
  \citenamefont {Berrington}, \citenamefont {Ahlefeldt},\ and\ \citenamefont
  {Longdell}}]{everts2019microwave}%
  \BibitemOpen
  \bibfield  {author} {\bibinfo {author} {\bibfnamefont {J.~R.}\ \bibnamefont
  {Everts}}, \bibinfo {author} {\bibfnamefont {M.~C.}\ \bibnamefont
  {Berrington}}, \bibinfo {author} {\bibfnamefont {R.~L.}\ \bibnamefont
  {Ahlefeldt}},\ and\ \bibinfo {author} {\bibfnamefont {J.~J.}\ \bibnamefont
  {Longdell}},\ }\bibfield  {title} {\bibinfo {title} {Microwave to optical
  photon conversion via fully concentrated rare-earth-ion crystals},\
  }\href@noop {} {\bibfield  {journal} {\bibinfo  {journal} {Physical Review
  A}\ }\textbf {\bibinfo {volume} {99}},\ \bibinfo {pages} {063830} (\bibinfo
  {year} {2019})}\BibitemShut {NoStop}%
\bibitem [{\citenamefont {Afzelius}\ \emph {et~al.}(2013)\citenamefont
  {Afzelius}, \citenamefont {Sangouard}, \citenamefont {Johansson},
  \citenamefont {Staudt},\ and\ \citenamefont {Wilson}}]{afzelius2013proposal}%
  \BibitemOpen
  \bibfield  {author} {\bibinfo {author} {\bibfnamefont {M.}~\bibnamefont
  {Afzelius}}, \bibinfo {author} {\bibfnamefont {N.}~\bibnamefont {Sangouard}},
  \bibinfo {author} {\bibfnamefont {G.}~\bibnamefont {Johansson}}, \bibinfo
  {author} {\bibfnamefont {M.}~\bibnamefont {Staudt}},\ and\ \bibinfo {author}
  {\bibfnamefont {C.}~\bibnamefont {Wilson}},\ }\bibfield  {title} {\bibinfo
  {title} {Proposal for a coherent quantum memory for propagating microwave
  photons},\ }\href@noop {} {\bibfield  {journal} {\bibinfo  {journal} {New
  Journal of Physics}\ }\textbf {\bibinfo {volume} {15}},\ \bibinfo {pages}
  {065008} (\bibinfo {year} {2013})}\BibitemShut {NoStop}%
\bibitem [{\citenamefont {Tkal{\v{c}}ec}\ \emph {et~al.}(2014)\citenamefont
  {Tkal{\v{c}}ec}, \citenamefont {Probst}, \citenamefont {Rieger},
  \citenamefont {Rotzinger}, \citenamefont {W{\"u}nsch}, \citenamefont
  {Kukharchyk}, \citenamefont {Wieck}, \citenamefont {Siegel}, \citenamefont
  {Ustinov},\ and\ \citenamefont {Bushev}}]{tkalvcec2014strong}%
  \BibitemOpen
  \bibfield  {author} {\bibinfo {author} {\bibfnamefont {A.}~\bibnamefont
  {Tkal{\v{c}}ec}}, \bibinfo {author} {\bibfnamefont {S.}~\bibnamefont
  {Probst}}, \bibinfo {author} {\bibfnamefont {D.}~\bibnamefont {Rieger}},
  \bibinfo {author} {\bibfnamefont {H.}~\bibnamefont {Rotzinger}}, \bibinfo
  {author} {\bibfnamefont {S.}~\bibnamefont {W{\"u}nsch}}, \bibinfo {author}
  {\bibfnamefont {N.}~\bibnamefont {Kukharchyk}}, \bibinfo {author}
  {\bibfnamefont {A.}~\bibnamefont {Wieck}}, \bibinfo {author} {\bibfnamefont
  {M.}~\bibnamefont {Siegel}}, \bibinfo {author} {\bibfnamefont
  {A.}~\bibnamefont {Ustinov}},\ and\ \bibinfo {author} {\bibfnamefont
  {P.}~\bibnamefont {Bushev}},\ }\bibfield  {title} {\bibinfo {title} {Strong
  coupling of an {{\rm Er$^{3+}$}-doped {\rm YAlO$_3$}} crystal to a
  superconducting resonator},\ }\href@noop {} {\bibfield  {journal} {\bibinfo
  {journal} {Physical Review B}\ }\textbf {\bibinfo {volume} {90}},\ \bibinfo
  {pages} {075112} (\bibinfo {year} {2014})}\BibitemShut {NoStop}%
\bibitem [{\citenamefont {Businger}\ \emph {et~al.}(2020)\citenamefont
  {Businger}, \citenamefont {Tiranov}, \citenamefont {Kaczmarek}, \citenamefont
  {Welinski}, \citenamefont {Zhang}, \citenamefont {Ferrier}, \citenamefont
  {Goldner},\ and\ \citenamefont {Afzelius}}]{businger2020optical}%
  \BibitemOpen
  \bibfield  {author} {\bibinfo {author} {\bibfnamefont {M.}~\bibnamefont
  {Businger}}, \bibinfo {author} {\bibfnamefont {A.}~\bibnamefont {Tiranov}},
  \bibinfo {author} {\bibfnamefont {K.~T.}\ \bibnamefont {Kaczmarek}}, \bibinfo
  {author} {\bibfnamefont {S.}~\bibnamefont {Welinski}}, \bibinfo {author}
  {\bibfnamefont {Z.}~\bibnamefont {Zhang}}, \bibinfo {author} {\bibfnamefont
  {A.}~\bibnamefont {Ferrier}}, \bibinfo {author} {\bibfnamefont
  {P.}~\bibnamefont {Goldner}},\ and\ \bibinfo {author} {\bibfnamefont
  {M.}~\bibnamefont {Afzelius}},\ }\bibfield  {title} {\bibinfo {title}
  {Optical spin-wave storage in a solid-state hybridized electron-nuclear spin
  ensemble},\ }\href@noop {} {\bibfield  {journal} {\bibinfo  {journal}
  {Physical Review Letters}\ }\textbf {\bibinfo {volume} {124}},\ \bibinfo
  {pages} {053606} (\bibinfo {year} {2020})}\BibitemShut {NoStop}%
\bibitem [{\citenamefont {Morton}\ \emph {et~al.}(2008)\citenamefont {Morton},
  \citenamefont {Tyryshkin}, \citenamefont {Brown}, \citenamefont {Shankar},
  \citenamefont {Lovett}, \citenamefont {Ardavan}, \citenamefont {Schenkel},
  \citenamefont {Haller}, \citenamefont {Ager},\ and\ \citenamefont
  {Lyon}}]{morton2008solid}%
  \BibitemOpen
  \bibfield  {author} {\bibinfo {author} {\bibfnamefont {J.~J.}\ \bibnamefont
  {Morton}}, \bibinfo {author} {\bibfnamefont {A.~M.}\ \bibnamefont
  {Tyryshkin}}, \bibinfo {author} {\bibfnamefont {R.~M.}\ \bibnamefont
  {Brown}}, \bibinfo {author} {\bibfnamefont {S.}~\bibnamefont {Shankar}},
  \bibinfo {author} {\bibfnamefont {B.~W.}\ \bibnamefont {Lovett}}, \bibinfo
  {author} {\bibfnamefont {A.}~\bibnamefont {Ardavan}}, \bibinfo {author}
  {\bibfnamefont {T.}~\bibnamefont {Schenkel}}, \bibinfo {author}
  {\bibfnamefont {E.~E.}\ \bibnamefont {Haller}}, \bibinfo {author}
  {\bibfnamefont {J.~W.}\ \bibnamefont {Ager}},\ and\ \bibinfo {author}
  {\bibfnamefont {S.~A.}\ \bibnamefont {Lyon}},\ }\bibfield  {title} {\bibinfo
  {title} {Solid-state quantum memory using the ${}^{31}p$ nuclear spin},\
  }\href@noop {} {\bibfield  {journal} {\bibinfo  {journal} {Nature}\ }\textbf
  {\bibinfo {volume} {455}},\ \bibinfo {pages} {1085} (\bibinfo {year}
  {2008})}\BibitemShut {NoStop}%
\bibitem [{\citenamefont {Rakonjac}\ \emph {et~al.}(2020)\citenamefont
  {Rakonjac}, \citenamefont {Chen}, \citenamefont {Horvath},\ and\
  \citenamefont {Longdell}}]{rakonjac2020long}%
  \BibitemOpen
  \bibfield  {author} {\bibinfo {author} {\bibfnamefont {J.~V.}\ \bibnamefont
  {Rakonjac}}, \bibinfo {author} {\bibfnamefont {Y.-H.}\ \bibnamefont {Chen}},
  \bibinfo {author} {\bibfnamefont {S.~P.}\ \bibnamefont {Horvath}},\ and\
  \bibinfo {author} {\bibfnamefont {J.~J.}\ \bibnamefont {Longdell}},\
  }\bibfield  {title} {\bibinfo {title} {{Long spin coherence times in the
  ground state and in an optically excited state of
  ${}^{167}$Er$^{3+}:$Y$_{2}$SiO$_{5}$ at zero magnetic field}},\ }\href@noop
  {} {\bibfield  {journal} {\bibinfo  {journal} {Physical Review B}\ }\textbf
  {\bibinfo {volume} {101}},\ \bibinfo {pages} {184430} (\bibinfo {year}
  {2020})}\BibitemShut {NoStop}%
\bibitem [{\citenamefont {B{\"o}ttger}\ \emph {et~al.}(2006)\citenamefont
  {B{\"o}ttger}, \citenamefont {Thiel}, \citenamefont {Sun},\ and\
  \citenamefont {Cone}}]{bottger2006optical}%
  \BibitemOpen
  \bibfield  {author} {\bibinfo {author} {\bibfnamefont {T.}~\bibnamefont
  {B{\"o}ttger}}, \bibinfo {author} {\bibfnamefont {C.}~\bibnamefont {Thiel}},
  \bibinfo {author} {\bibfnamefont {Y.}~\bibnamefont {Sun}},\ and\ \bibinfo
  {author} {\bibfnamefont {R.}~\bibnamefont {Cone}},\ }\bibfield  {title}
  {\bibinfo {title} {{Optical decoherence and spectral diffusion at 1.5 $\mu$m
  in ${}^{167}$Er$^{3+}:$Y$_{2}$SiO$_{5}$ versus magnetic field, temperature,
  and Er$^{3+}$ concentration}},\ }\href@noop {} {\bibfield  {journal}
  {\bibinfo  {journal} {Physical Review B}\ }\textbf {\bibinfo {volume} {73}},\
  \bibinfo {pages} {075101} (\bibinfo {year} {2006})}\BibitemShut {NoStop}%
\bibitem [{\citenamefont {Klauder}\ and\ \citenamefont
  {Anderson}(1962)}]{klauder1962spectral}%
  \BibitemOpen
  \bibfield  {author} {\bibinfo {author} {\bibfnamefont {J.}~\bibnamefont
  {Klauder}}\ and\ \bibinfo {author} {\bibfnamefont {P.}~\bibnamefont
  {Anderson}},\ }\bibfield  {title} {\bibinfo {title} {Spectral diffusion decay
  in spin resonance experiments},\ }\href@noop {} {\bibfield  {journal}
  {\bibinfo  {journal} {Physical Review}\ }\textbf {\bibinfo {volume} {125}},\
  \bibinfo {pages} {912} (\bibinfo {year} {1962})}\BibitemShut {NoStop}%
\bibitem [{\citenamefont {Li}\ \emph {et~al.}(2020)\citenamefont {Li},
  \citenamefont {Liu}, \citenamefont {Zhou}, \citenamefont {Liu}, \citenamefont
  {Tu}, \citenamefont {Yang}, \citenamefont {Li}, \citenamefont {Ma},
  \citenamefont {Hu}, \citenamefont {Liang} \emph {et~al.}}]{li2020hyperfine}%
  \BibitemOpen
  \bibfield  {author} {\bibinfo {author} {\bibfnamefont {P.-Y.}\ \bibnamefont
  {Li}}, \bibinfo {author} {\bibfnamefont {C.}~\bibnamefont {Liu}}, \bibinfo
  {author} {\bibfnamefont {Z.-Q.}\ \bibnamefont {Zhou}}, \bibinfo {author}
  {\bibfnamefont {X.}~\bibnamefont {Liu}}, \bibinfo {author} {\bibfnamefont
  {T.}~\bibnamefont {Tu}}, \bibinfo {author} {\bibfnamefont {T.-S.}\
  \bibnamefont {Yang}}, \bibinfo {author} {\bibfnamefont {Z.-F.}\ \bibnamefont
  {Li}}, \bibinfo {author} {\bibfnamefont {Y.}~\bibnamefont {Ma}}, \bibinfo
  {author} {\bibfnamefont {J.}~\bibnamefont {Hu}}, \bibinfo {author}
  {\bibfnamefont {P.-J.}\ \bibnamefont {Liang}}, \emph {et~al.},\ }\bibfield
  {title} {\bibinfo {title} {Hyperfine structure and coherent dynamics of
  rare-earth spins explored with electron-nuclear double resonance at subkelvin
  temperatures},\ }\href@noop {} {\bibfield  {journal} {\bibinfo  {journal}
  {Physical Review Applied}\ }\textbf {\bibinfo {volume} {13}},\ \bibinfo
  {pages} {024080} (\bibinfo {year} {2020})}\BibitemShut {NoStop}%
\bibitem [{\citenamefont {Kukharchyk}\ \emph {et~al.}(2019)\citenamefont
  {Kukharchyk}, \citenamefont {Sholokhov}, \citenamefont {Kalachev},\ and\
  \citenamefont {Bushev}}]{kukharchyk2019enhancement}%
  \BibitemOpen
  \bibfield  {author} {\bibinfo {author} {\bibfnamefont {N.}~\bibnamefont
  {Kukharchyk}}, \bibinfo {author} {\bibfnamefont {D.}~\bibnamefont
  {Sholokhov}}, \bibinfo {author} {\bibfnamefont {A.}~\bibnamefont
  {Kalachev}},\ and\ \bibinfo {author} {\bibfnamefont {P.}~\bibnamefont
  {Bushev}},\ }\bibfield  {title} {\bibinfo {title} {{Enhancement of optical
  coherence in ${}^{167}$Er$^{3+}:$Y$_{2}$SiO$_{5}$ crystal at sub-Kelvin
  temperatures}},\ }\href@noop {} {\bibfield  {journal} {\bibinfo  {journal}
  {arXiv preprint arXiv:1910.03096}\ } (\bibinfo {year} {2019})}\BibitemShut
  {NoStop}%
\bibitem [{\citenamefont {Feher}\ and\ \citenamefont
  {Gere}(1959)}]{feher1959electron}%
  \BibitemOpen
  \bibfield  {author} {\bibinfo {author} {\bibfnamefont {G.}~\bibnamefont
  {Feher}}\ and\ \bibinfo {author} {\bibfnamefont {E.}~\bibnamefont {Gere}},\
  }\bibfield  {title} {\bibinfo {title} {Electron spin resonance experiments on
  donors in silicon. ii. electron spin relaxation effects},\ }\href@noop {}
  {\bibfield  {journal} {\bibinfo  {journal} {Physical Review}\ }\textbf
  {\bibinfo {volume} {114}},\ \bibinfo {pages} {1245} (\bibinfo {year}
  {1959})}\BibitemShut {NoStop}%
\bibitem [{\citenamefont {Wolfowicz}\ \emph {et~al.}(2015)\citenamefont
  {Wolfowicz}, \citenamefont {Maier-Flaig}, \citenamefont {Marino},
  \citenamefont {Ferrier}, \citenamefont {Vezin}, \citenamefont {Morton},\ and\
  \citenamefont {Goldner}}]{wolfowicz2015coherent}%
  \BibitemOpen
  \bibfield  {author} {\bibinfo {author} {\bibfnamefont {G.}~\bibnamefont
  {Wolfowicz}}, \bibinfo {author} {\bibfnamefont {H.}~\bibnamefont
  {Maier-Flaig}}, \bibinfo {author} {\bibfnamefont {R.}~\bibnamefont {Marino}},
  \bibinfo {author} {\bibfnamefont {A.}~\bibnamefont {Ferrier}}, \bibinfo
  {author} {\bibfnamefont {H.}~\bibnamefont {Vezin}}, \bibinfo {author}
  {\bibfnamefont {J.~J.}\ \bibnamefont {Morton}},\ and\ \bibinfo {author}
  {\bibfnamefont {P.}~\bibnamefont {Goldner}},\ }\bibfield  {title} {\bibinfo
  {title} {Coherent storage of microwave excitations in rare-earth nuclear
  spins},\ }\href@noop {} {\bibfield  {journal} {\bibinfo  {journal} {Physical
  Review Letters}\ }\textbf {\bibinfo {volume} {114}},\ \bibinfo {pages}
  {170503} (\bibinfo {year} {2015})}\BibitemShut {NoStop}%
\bibitem [{\citenamefont {Kindem}\ \emph {et~al.}(2018)\citenamefont {Kindem},
  \citenamefont {Bartholomew}, \citenamefont {Woodburn}, \citenamefont {Zhong},
  \citenamefont {Craiciu}, \citenamefont {Cone}, \citenamefont {Thiel},\ and\
  \citenamefont {Faraon}}]{kindem2018characterization}%
  \BibitemOpen
  \bibfield  {author} {\bibinfo {author} {\bibfnamefont {J.~M.}\ \bibnamefont
  {Kindem}}, \bibinfo {author} {\bibfnamefont {J.~G.}\ \bibnamefont
  {Bartholomew}}, \bibinfo {author} {\bibfnamefont {P.~J.}\ \bibnamefont
  {Woodburn}}, \bibinfo {author} {\bibfnamefont {T.}~\bibnamefont {Zhong}},
  \bibinfo {author} {\bibfnamefont {I.}~\bibnamefont {Craiciu}}, \bibinfo
  {author} {\bibfnamefont {R.~L.}\ \bibnamefont {Cone}}, \bibinfo {author}
  {\bibfnamefont {C.~W.}\ \bibnamefont {Thiel}},\ and\ \bibinfo {author}
  {\bibfnamefont {A.}~\bibnamefont {Faraon}},\ }\bibfield  {title} {\bibinfo
  {title} {{Characterization of ${}^{171}$Yb$^{3+}:$YVO$_{4}$ for photonic
  quantum technologies}},\ }\href@noop {} {\bibfield  {journal} {\bibinfo
  {journal} {Physical Review B}\ }\textbf {\bibinfo {volume} {98}},\ \bibinfo
  {pages} {024404} (\bibinfo {year} {2018})}\BibitemShut {NoStop}%
\bibitem [{\citenamefont {Sun}\ \emph {et~al.}(2008)\citenamefont {Sun},
  \citenamefont {B{\"o}ttger}, \citenamefont {Thiel},\ and\ \citenamefont
  {Cone}}]{sun2008magnetic}%
  \BibitemOpen
  \bibfield  {author} {\bibinfo {author} {\bibfnamefont {Y.}~\bibnamefont
  {Sun}}, \bibinfo {author} {\bibfnamefont {T.}~\bibnamefont {B{\"o}ttger}},
  \bibinfo {author} {\bibfnamefont {C.}~\bibnamefont {Thiel}},\ and\ \bibinfo
  {author} {\bibfnamefont {R.}~\bibnamefont {Cone}},\ }\bibfield  {title}
  {\bibinfo {title} {{Magnetic g tensors for the ${}^{4}I_{15/2}$ and
  ${}^{4}I_{13/2}$ states of Er$^{3+}:$Y$_{2}$SiO$_{5}$}},\ }\href@noop {}
  {\bibfield  {journal} {\bibinfo  {journal} {Physical Review B}\ }\textbf
  {\bibinfo {volume} {77}},\ \bibinfo {pages} {085124} (\bibinfo {year}
  {2008})}\BibitemShut {NoStop}%
\bibitem [{\citenamefont {Thiel}\ \emph {et~al.}(2011)\citenamefont {Thiel},
  \citenamefont {B{\"o}ttger},\ and\ \citenamefont {Cone}}]{thiel2011rare}%
  \BibitemOpen
  \bibfield  {author} {\bibinfo {author} {\bibfnamefont {C.}~\bibnamefont
  {Thiel}}, \bibinfo {author} {\bibfnamefont {T.}~\bibnamefont {B{\"o}ttger}},\
  and\ \bibinfo {author} {\bibfnamefont {R.}~\bibnamefont {Cone}},\ }\bibfield
  {title} {\bibinfo {title} {Rare-earth-doped materials for applications in
  quantum information storage and signal processing},\ }\href@noop {}
  {\bibfield  {journal} {\bibinfo  {journal} {Journal of luminescence}\
  }\textbf {\bibinfo {volume} {131}},\ \bibinfo {pages} {353} (\bibinfo {year}
  {2011})}\BibitemShut {NoStop}%
\bibitem [{\citenamefont {Sigillito}\ \emph {et~al.}(2017)\citenamefont
  {Sigillito}, \citenamefont {Tyryshkin}, \citenamefont {Schenkel},
  \citenamefont {Houck},\ and\ \citenamefont
  {Lyon}}]{sigillito2017all-electric}%
  \BibitemOpen
  \bibfield  {author} {\bibinfo {author} {\bibfnamefont {A.~J.}\ \bibnamefont
  {Sigillito}}, \bibinfo {author} {\bibfnamefont {A.~M.}\ \bibnamefont
  {Tyryshkin}}, \bibinfo {author} {\bibfnamefont {T.}~\bibnamefont {Schenkel}},
  \bibinfo {author} {\bibfnamefont {A.}~\bibnamefont {Houck}},\ and\ \bibinfo
  {author} {\bibfnamefont {S.~A.}\ \bibnamefont {Lyon}},\ }\bibfield  {title}
  {\bibinfo {title} {All-electric control of donor nuclear spin qubits in
  silicon},\ }\href@noop {} {\bibfield  {journal} {\bibinfo  {journal} {Nature
  Nanotechnology}\ }\textbf {\bibinfo {volume} {12}},\ \bibinfo {pages} {958}
  (\bibinfo {year} {2017})}\BibitemShut {NoStop}%
\bibitem [{\citenamefont {Guillot-No{\"e}l}\ \emph {et~al.}(2006)\citenamefont
  {Guillot-No{\"e}l}, \citenamefont {Goldner}, \citenamefont {Le~Du},
  \citenamefont {Baldit}, \citenamefont {Monnier},\ and\ \citenamefont
  {Bencheikh}}]{guillot2006hyperfine}%
  \BibitemOpen
  \bibfield  {author} {\bibinfo {author} {\bibfnamefont {O.}~\bibnamefont
  {Guillot-No{\"e}l}}, \bibinfo {author} {\bibfnamefont {P.}~\bibnamefont
  {Goldner}}, \bibinfo {author} {\bibfnamefont {Y.}~\bibnamefont {Le~Du}},
  \bibinfo {author} {\bibfnamefont {E.}~\bibnamefont {Baldit}}, \bibinfo
  {author} {\bibfnamefont {P.}~\bibnamefont {Monnier}},\ and\ \bibinfo {author}
  {\bibfnamefont {K.}~\bibnamefont {Bencheikh}},\ }\bibfield  {title} {\bibinfo
  {title} {{Hyperfine interaction of Er$^{3+}$ ions in Y$_{2}$SiO$_{5}$: An
  electron paramagnetic resonance spectroscopy study}},\ }\href@noop {}
  {\bibfield  {journal} {\bibinfo  {journal} {Physical Review B}\ }\textbf
  {\bibinfo {volume} {74}},\ \bibinfo {pages} {214409} (\bibinfo {year}
  {2006})}\BibitemShut {NoStop}%
\bibitem [{\citenamefont {Abragam}\ and\ \citenamefont
  {Bleaney}(2012)}]{abragam2012electron}%
  \BibitemOpen
  \bibfield  {author} {\bibinfo {author} {\bibfnamefont {A.}~\bibnamefont
  {Abragam}}\ and\ \bibinfo {author} {\bibfnamefont {B.}~\bibnamefont
  {Bleaney}},\ }\href@noop {} {\emph {\bibinfo {title} {Electron paramagnetic
  resonance of transition ions}}}\ (\bibinfo  {publisher} {OUP Oxford},\
  \bibinfo {year} {2012})\BibitemShut {NoStop}%
\bibitem [{\citenamefont {Chen}\ \emph {et~al.}(2018)\citenamefont {Chen},
  \citenamefont {Fernandez-Gonzalvo}, \citenamefont {Horvath}, \citenamefont
  {Rakonjac},\ and\ \citenamefont {Longdell}}]{chen2018hyperfine}%
  \BibitemOpen
  \bibfield  {author} {\bibinfo {author} {\bibfnamefont {Y.-H.}\ \bibnamefont
  {Chen}}, \bibinfo {author} {\bibfnamefont {X.}~\bibnamefont
  {Fernandez-Gonzalvo}}, \bibinfo {author} {\bibfnamefont {S.~P.}\ \bibnamefont
  {Horvath}}, \bibinfo {author} {\bibfnamefont {J.~V.}\ \bibnamefont
  {Rakonjac}},\ and\ \bibinfo {author} {\bibfnamefont {J.~J.}\ \bibnamefont
  {Longdell}},\ }\bibfield  {title} {\bibinfo {title} {{Hyperfine interactions
  of Er$^{3+}$ ions in Y$_{2}$SiO$_{5}$: Electron paramagnetic resonance in a
  tunable microwave cavity}},\ }\href@noop {} {\bibfield  {journal} {\bibinfo
  {journal} {Physical Review B}\ }\textbf {\bibinfo {volume} {97}},\ \bibinfo
  {pages} {024419} (\bibinfo {year} {2018})}\BibitemShut {NoStop}%
\bibitem [{\citenamefont {Schweiger}\ and\ \citenamefont
  {Jeschke}(2001)}]{schweiger2001principles}%
  \BibitemOpen
  \bibfield  {author} {\bibinfo {author} {\bibfnamefont {A.}~\bibnamefont
  {Schweiger}}\ and\ \bibinfo {author} {\bibfnamefont {G.}~\bibnamefont
  {Jeschke}},\ }\href@noop {} {\emph {\bibinfo {title} {Principles of pulse
  electron paramagnetic resonance}}}\ (\bibinfo  {publisher} {Oxford University
  Press on Demand},\ \bibinfo {year} {2001})\BibitemShut {NoStop}%
\bibitem [{\citenamefont {Orbach}(1961)}]{orbach1961spin}%
  \BibitemOpen
  \bibfield  {author} {\bibinfo {author} {\bibfnamefont {R.}~\bibnamefont
  {Orbach}},\ }\bibfield  {title} {\bibinfo {title} {Spin-lattice relaxation in
  rare-earth salts},\ }\href@noop {} {\bibfield  {journal} {\bibinfo  {journal}
  {Proceedings of the Royal Society of London. Series A. Mathematical and
  Physical Sciences}\ }\textbf {\bibinfo {volume} {264}},\ \bibinfo {pages}
  {458} (\bibinfo {year} {1961})}\BibitemShut {NoStop}%
\bibitem [{\citenamefont {Budoyo}\ \emph {et~al.}(2018)\citenamefont {Budoyo},
  \citenamefont {Kakuyanagi}, \citenamefont {Toida}, \citenamefont {Matsuzaki},
  \citenamefont {Munro}, \citenamefont {Yamaguchi},\ and\ \citenamefont
  {Saito}}]{budoyo2018phonon}%
  \BibitemOpen
  \bibfield  {author} {\bibinfo {author} {\bibfnamefont {R.~P.}\ \bibnamefont
  {Budoyo}}, \bibinfo {author} {\bibfnamefont {K.}~\bibnamefont {Kakuyanagi}},
  \bibinfo {author} {\bibfnamefont {H.}~\bibnamefont {Toida}}, \bibinfo
  {author} {\bibfnamefont {Y.}~\bibnamefont {Matsuzaki}}, \bibinfo {author}
  {\bibfnamefont {W.~J.}\ \bibnamefont {Munro}}, \bibinfo {author}
  {\bibfnamefont {H.}~\bibnamefont {Yamaguchi}},\ and\ \bibinfo {author}
  {\bibfnamefont {S.}~\bibnamefont {Saito}},\ }\bibfield  {title} {\bibinfo
  {title} {Phonon-bottlenecked spin relaxation of {Er$^{3+}$:Y$_{2}$SiO$_{5}$}
  at sub-kelvin temperatures},\ }\href@noop {} {\bibfield  {journal} {\bibinfo
  {journal} {Applied Physics Express}\ }\textbf {\bibinfo {volume} {11}},\
  \bibinfo {pages} {043002} (\bibinfo {year} {2018})}\BibitemShut {NoStop}%
\bibitem [{\citenamefont {Mims}(1968)}]{mims1968phase}%
  \BibitemOpen
  \bibfield  {author} {\bibinfo {author} {\bibfnamefont {W.}~\bibnamefont
  {Mims}},\ }\bibfield  {title} {\bibinfo {title} {{Phase memory in electron
  spin echoes, lattice relaxation effects in CaWO$_{4}:$Er, Ce, Mn}},\
  }\href@noop {} {\bibfield  {journal} {\bibinfo  {journal} {Physical Review}\
  }\textbf {\bibinfo {volume} {168}},\ \bibinfo {pages} {370} (\bibinfo {year}
  {1968})}\BibitemShut {NoStop}%
\bibitem [{\citenamefont {Tyryshkin}\ \emph {et~al.}(2012)\citenamefont
  {Tyryshkin}, \citenamefont {Tojo}, \citenamefont {Morton}, \citenamefont
  {Riemann}, \citenamefont {Abrosimov}, \citenamefont {Becker}, \citenamefont
  {Pohl}, \citenamefont {Schenkel}, \citenamefont {Thewalt}, \citenamefont
  {Itoh} \emph {et~al.}}]{tyryshkin2012electron}%
  \BibitemOpen
  \bibfield  {author} {\bibinfo {author} {\bibfnamefont {A.~M.}\ \bibnamefont
  {Tyryshkin}}, \bibinfo {author} {\bibfnamefont {S.}~\bibnamefont {Tojo}},
  \bibinfo {author} {\bibfnamefont {J.~J.}\ \bibnamefont {Morton}}, \bibinfo
  {author} {\bibfnamefont {H.}~\bibnamefont {Riemann}}, \bibinfo {author}
  {\bibfnamefont {N.~V.}\ \bibnamefont {Abrosimov}}, \bibinfo {author}
  {\bibfnamefont {P.}~\bibnamefont {Becker}}, \bibinfo {author} {\bibfnamefont
  {H.-J.}\ \bibnamefont {Pohl}}, \bibinfo {author} {\bibfnamefont
  {T.}~\bibnamefont {Schenkel}}, \bibinfo {author} {\bibfnamefont {M.~L.}\
  \bibnamefont {Thewalt}}, \bibinfo {author} {\bibfnamefont {K.~M.}\
  \bibnamefont {Itoh}}, \emph {et~al.},\ }\bibfield  {title} {\bibinfo {title}
  {Electron spin coherence exceeding seconds in high-purity silicon},\
  }\href@noop {} {\bibfield  {journal} {\bibinfo  {journal} {Nature materials}\
  }\textbf {\bibinfo {volume} {11}},\ \bibinfo {pages} {143} (\bibinfo {year}
  {2012})}\BibitemShut {NoStop}%
\bibitem [{\citenamefont {Kutter}\ \emph {et~al.}(1995)\citenamefont {Kutter},
  \citenamefont {Moll}, \citenamefont {Van~Tol}, \citenamefont {Zuckermann},
  \citenamefont {Maan},\ and\ \citenamefont {Wyder}}]{kutter1995electron}%
  \BibitemOpen
  \bibfield  {author} {\bibinfo {author} {\bibfnamefont {C.}~\bibnamefont
  {Kutter}}, \bibinfo {author} {\bibfnamefont {H.}~\bibnamefont {Moll}},
  \bibinfo {author} {\bibfnamefont {J.}~\bibnamefont {Van~Tol}}, \bibinfo
  {author} {\bibfnamefont {H.}~\bibnamefont {Zuckermann}}, \bibinfo {author}
  {\bibfnamefont {J.}~\bibnamefont {Maan}},\ and\ \bibinfo {author}
  {\bibfnamefont {P.}~\bibnamefont {Wyder}},\ }\bibfield  {title} {\bibinfo
  {title} {Electron-spin echoes at 604 ghz using far infrared lasers},\
  }\href@noop {} {\bibfield  {journal} {\bibinfo  {journal} {Physical Review
  Letters}\ }\textbf {\bibinfo {volume} {74}},\ \bibinfo {pages} {2925}
  (\bibinfo {year} {1995})}\BibitemShut {NoStop}%
\bibitem [{\citenamefont {Dzuba}\ and\ \citenamefont
  {Kawamori}(1996)}]{dzuba1996selective}%
  \BibitemOpen
  \bibfield  {author} {\bibinfo {author} {\bibfnamefont {S.~A.}\ \bibnamefont
  {Dzuba}}\ and\ \bibinfo {author} {\bibfnamefont {A.}~\bibnamefont
  {Kawamori}},\ }\bibfield  {title} {\bibinfo {title} {Selective hole burning
  in epr: Spectral diffusion and dipolar broadening},\ }\href@noop {}
  {\bibfield  {journal} {\bibinfo  {journal} {Concepts in Magnetic Resonance}\
  }\textbf {\bibinfo {volume} {8}},\ \bibinfo {pages} {49} (\bibinfo {year}
  {1996})}\BibitemShut {NoStop}%
\bibitem [{\citenamefont {Probst}\ \emph {et~al.}(2014)\citenamefont {Probst},
  \citenamefont {Kukharchyk}, \citenamefont {Rotzinger}, \citenamefont
  {Tkal{\v{c}}ec}, \citenamefont {W{\"u}nsch}, \citenamefont {Wieck},
  \citenamefont {Siegel}, \citenamefont {Ustinov},\ and\ \citenamefont
  {Bushev}}]{probst2014hybrid}%
  \BibitemOpen
  \bibfield  {author} {\bibinfo {author} {\bibfnamefont {S.}~\bibnamefont
  {Probst}}, \bibinfo {author} {\bibfnamefont {N.}~\bibnamefont {Kukharchyk}},
  \bibinfo {author} {\bibfnamefont {H.}~\bibnamefont {Rotzinger}}, \bibinfo
  {author} {\bibfnamefont {A.}~\bibnamefont {Tkal{\v{c}}ec}}, \bibinfo {author}
  {\bibfnamefont {S.}~\bibnamefont {W{\"u}nsch}}, \bibinfo {author}
  {\bibfnamefont {A.}~\bibnamefont {Wieck}}, \bibinfo {author} {\bibfnamefont
  {M.}~\bibnamefont {Siegel}}, \bibinfo {author} {\bibfnamefont
  {A.}~\bibnamefont {Ustinov}},\ and\ \bibinfo {author} {\bibfnamefont
  {P.}~\bibnamefont {Bushev}},\ }\bibfield  {title} {\bibinfo {title} {Hybrid
  quantum circuit with implanted erbium ions},\ }\href@noop {} {\bibfield
  {journal} {\bibinfo  {journal} {Applied Physics Letters}\ }\textbf {\bibinfo
  {volume} {105}},\ \bibinfo {pages} {162404} (\bibinfo {year}
  {2014})}\BibitemShut {NoStop}%
\bibitem [{\citenamefont {Craiciu}\ \emph {et~al.}(2021)\citenamefont
  {Craiciu}, \citenamefont {Lei}, \citenamefont {Rochman}, \citenamefont
  {Bartholomew},\ and\ \citenamefont {Faraon}}]{craiciu2021multifunctional}%
  \BibitemOpen
  \bibfield  {author} {\bibinfo {author} {\bibfnamefont {I.}~\bibnamefont
  {Craiciu}}, \bibinfo {author} {\bibfnamefont {M.}~\bibnamefont {Lei}},
  \bibinfo {author} {\bibfnamefont {J.}~\bibnamefont {Rochman}}, \bibinfo
  {author} {\bibfnamefont {J.~G.}\ \bibnamefont {Bartholomew}},\ and\ \bibinfo
  {author} {\bibfnamefont {A.}~\bibnamefont {Faraon}},\ }\bibfield  {title}
  {\bibinfo {title} {Multifunctional on-chip storage at telecommunication
  wavelength for quantum networks},\ }\href@noop {} {\bibfield  {journal}
  {\bibinfo  {journal} {Optica}\ }\textbf {\bibinfo {volume} {8}},\ \bibinfo
  {pages} {114} (\bibinfo {year} {2021})}\BibitemShut {NoStop}%
\bibitem [{\citenamefont {Zaripov}\ \emph {et~al.}(2013)\citenamefont
  {Zaripov}, \citenamefont {Vavilova}, \citenamefont {Miluykov}, \citenamefont
  {Bezkishko}, \citenamefont {Sinyashin}, \citenamefont {Salikhov},
  \citenamefont {Kataev},\ and\ \citenamefont
  {B{\"u}chner}}]{zaripov2013boosting}%
  \BibitemOpen
  \bibfield  {author} {\bibinfo {author} {\bibfnamefont {R.}~\bibnamefont
  {Zaripov}}, \bibinfo {author} {\bibfnamefont {E.}~\bibnamefont {Vavilova}},
  \bibinfo {author} {\bibfnamefont {V.}~\bibnamefont {Miluykov}}, \bibinfo
  {author} {\bibfnamefont {I.}~\bibnamefont {Bezkishko}}, \bibinfo {author}
  {\bibfnamefont {O.}~\bibnamefont {Sinyashin}}, \bibinfo {author}
  {\bibfnamefont {K.}~\bibnamefont {Salikhov}}, \bibinfo {author}
  {\bibfnamefont {V.}~\bibnamefont {Kataev}},\ and\ \bibinfo {author}
  {\bibfnamefont {B.}~\bibnamefont {B{\"u}chner}},\ }\bibfield  {title}
  {\bibinfo {title} {Boosting the electron spin coherence in binuclear mn
  complexes by multiple microwave pulses},\ }\href@noop {} {\bibfield
  {journal} {\bibinfo  {journal} {Physical Review B}\ }\textbf {\bibinfo
  {volume} {88}},\ \bibinfo {pages} {094418} (\bibinfo {year}
  {2013})}\BibitemShut {NoStop}%
\end{thebibliography}%
	
\end{document}